\documentclass{article}

\usepackage{PRIMEarxiv}
\usepackage{natbib}
\usepackage[utf8]{inputenc} 
\usepackage[T1]{fontenc}    

\usepackage{hyperref}       
\usepackage{url}            
\usepackage{booktabs}       
\usepackage{amsfonts}       
\usepackage{nicefrac}       
\usepackage{microtype}      
\usepackage{lipsum}
\usepackage{fancyhdr}       
\usepackage{graphicx}       
\usepackage{amsmath}
\usepackage{booktabs}
\usepackage{caption}
\usepackage{graphicx}
\usepackage{diagbox}
\usepackage{siunitx}
\usepackage{setspace}
\usepackage{subcaption}
\usepackage{amsthm}
\usepackage[table]{xcolor}
\usepackage{rotating}
\usepackage{xcolor}
\usepackage{footnote}
\usepackage{longtable}
\usepackage{enumitem} 
\usepackage{makecell} 
\usepackage{framed}
\usepackage{booktabs} 
\usepackage{array}
\usepackage[linesnumbered,lined,boxed,commentsnumbered]{algorithm2e} 
\usepackage{bbm}
\usepackage[linesnumbered,lined,boxed,commentsnumbered]{algorithm2e}
\DeclareMathOperator*{\argmin}{arg\,min}
\DeclareMathOperator*{\argmax}{arg\,max}

\theoremstyle{definition}

\definecolor{lightgray}{gray}{0.8}
\definecolor{darkgray}{gray}{0.9}
\definecolor{lgray}{rgb}{0.9, 0.9, 0.9}
\graphicspath{{media/}}     
\title{ BRBVS: An \texttt{R} Package for Bivariate Variable selection in Copula Survival Model(s) domain.
}

\author{
      Danilo Petti\thanks{Corresponding author.}\thanks{School of Mathematics, University of Essex, Wivenhoe Park, Colchester.\texttt{d.petti@essex.ac.uk}.}
       \and
       Marcella Niglio \thanks{Department of Statistics, University of Salerno, Via Giovanni Paolo II, 132, Fisciano. \texttt{mnigli@unisa.it}.}
       \and
       Marialuisa Restaino \thanks{Department of Statistics, University of Salerno, Via Giovanni Paolo II, 132, Fisciano. \texttt{mlrestaino@unisa.it}.}
}

\begin{document}
\maketitle

\begin{abstract}
\texttt{BRBVS}  is a publicly available \texttt{R} package on CRAN that implements the  algorithm proposed in \cite{Petti2024sub}.   The algorithm was developed as the first proposal of variable selection for the class of Bivariate Survival Copula Models originally proposed in \cite{Marra2020} and implemented in the \texttt{GJRM} package. The core of the \texttt{BRBVS} package is to implement and make available to practitioners variable selection algorithms for bivariate survival data affected by censoring, providing easy-to-use functions and graphical outputs. The idea behind the algorithm is almost general and may also be extended to different class of models.
\end{abstract}

\keywords{ Bivariate survival data \and Copula \and Mixed censoring scheme\and  Variable selection.}

\clearpage
\tableofcontents

\section{Motivation}\label{intro}

The growing opportunities of collecting, storing and sharing data has lead to build huge datasets with million of observations characterized by 
a large number of covariates that sometimes also exceed the number of units ($p \gg n$). In this case 
the selection of the most relevant and significant variables is crucial to enhance the quality of estimation, prediction, and interpretation within models \citep{fan2001variable}.

Over the past few decades, an extensive body of literature on variable selection methods has played a pivotal role across many fields, especially in the context of linear models \cite[see reviews by][]{fan2010overviewvarableselectio, desboulets2018review, HeinzeEtAl2018}. Traditional techniques widely applied for variable selection include forward selection, backward elimination, stepwise selection, and best-subset selection \citep{harrell2001}. Additionally, penalized variable selection methods are developed for linear regression \citep{tibshirani1996regression}, generalized linear models \citep{Friedman2010}, accelerated failure time models (AFT model) \citep{park2018aft}, Cox's proportional hazards and frailty models \citep{tibshirani1997lasso, fan2002variable, zou2008note, fan2001variable}, copula survival models \citep{Kwon2019copula}, and multivariate survival data \citep{cai2005variable}.

To address the disadvantages of penalization methods (e.g., computational efficiency, statistical accuracy, and algorithmic stability), screening procedures are proposed \citep{fan2008, Fan2009VS}  to contain the number of variables. These procedures rank variables in terms of their importance, measured by the association between the dependent variable and features.

Furthermore, variable selection algorithms can be improved by combining screening procedures with bootstrap and permutation tests, as proposed by \cite{Baranowski2020} in the regression domain. This procedure, called Ranking-Based Variable Selection (RBVS), identifies subsets of covariates that consistently appear to be important across subsamples extracted from the data.

 From a software perspective, several packages which deal with selecting the relevant features are available. To give some examples, the \texttt{R} package \texttt{CoxICPen} \citep{CoxICPen} allows for variable selection on Cox models affected by interval censoring. In the high-dimensional domain, the \texttt{highMLR} package in \texttt{R} \citep{highMLR} performs high-dimensional feature selection in the presence of survival outcomes.

Additional packages addressing variable selection in survival analysis include \texttt{glmnet} \citep{glmnet}, which implements penalized variable selection for linear and generalized linear models, and \texttt{SIS} \citep{SIS}, offering sure independence screening for generalized linear models and Cox proportional hazards models. Furthermore, the \texttt{R} package \texttt{penPHcure} \citep{penPHcure} provides variable selection procedures for semi-parametric proportional hazards cure models, accommodating time-varying covariates.

In the multivariate context with continuous outcomes, outside the survival domain, the \texttt{R} software offers the \texttt{mBvs} \citep{mBvs} and \texttt{MultiVarSel} \citep{MultiVarSel} packages. The former proposes Bayesian variable selection methods for data with multivariate responses and multiple covariates, while the latter performs variable selection in a multivariate linear model.

In other  programming languages for Data Analysts and Scientists, such as \texttt{Python} and \texttt{Julia}, variable selection methods are only marginally implemented, with survival data-based methods entirely lacking.

Methologies and software performing variable selection for bivariate survival functions are not available, leading to adapting univariate methods to a bi(multi)variate settings. The main risk is related to the increase in the false positive during the selection process. This happens because univariate methods: a) do not allow for the selection of covariates that maximize the dependency relationship between the two survival functions; b)  completely ignores the dependency structure that may exists between the two survival functions; c) are not configured to handle various censoring mechanisms.


Given this background, in this paper, we present the \texttt{BRBVS} package in \texttt{R} \citep{Petti2024Package}, which implements the Bivariate Variable Ranking Based Variable Selection algorithm (BRBVS) proposed in \cite{Petti2024sub}, extending the method introduced by \cite{Baranowski2020}, originally developed for linear models. The BRBVS algorithm is capable to handle two rankings, as copula survival contexts involve two potentially related survival functions influenced by censoring mechanisms.

From a computational standpoint, the BRBVS algorithm consists of two main steps:
1) \textit{Variable Ranking}, where bootstrap sampling based on a subsample of the dataset is used to construct two rankings through a relevance measure. Each ranking corresponds to one of the survival functions and has a cardinality equal to the number of covariates in the dataset;
2) \textit{Variable Selection}, in which two sets of relevant variables are defined, one for each margin. These sets are chosen from the covariates that most frequently appear in the top positions of the two rankings \textbf{obtained for each extracted subsample}.

In the current version of the package, only the class of Copula Link Based Survival Model(s)  is implemented. The extension to other classes of bivariate survival models may be considered in the future versions of the package, as the algorithm can be easily generalized, thanks to the flexibility in the analytical structure \citep{Petti2024sub}.  Furthermore, the measure introduced to rank the covariates is almost general and may also be considered to make variable selection for other classes of models.  The \texttt{BRBVS} package is available under the General Public License (GPL $\geq 3$) from the Comprehensive R Archive Network at \href{https://cran.r-project.org/web/packages/BRBVS/index.html}{https://cran.r-project.org/web/packages/BRBVS}.

The paper is organized as follows. In Section \ref{sec:Framework}, we provide an overview of the model framework and algorithm, beginning with the model framework. More specifically, Section \ref{Ch:modelformulation} offers an overview of the Copula Link-Based Survival Models, while Section \ref{sec:BRBVS} presents a detailed description of the algorithm. Section \ref{sec:packageDetails} is dedicated to describing the main functions implemented in the \texttt{BRBVS} package, showcasing their application with real-world data. In Section \ref{sec:simulation}, a Monte Carlo study gives evidence of the consistency of the algorithm. Finally, Section \ref{sec:summary} concludes the paper with a discussion.

\clearpage
\section{Theoretical Framework}
\label{sec:Framework}

In this section, we will recall the  theoretical framework  behind the Bivariate Variable Ranking Based Variable Selection Algorithm (BRBVS). The BRBVS method is the first algorithm that enables variable selection in bivariate survival contexts with censored data for the class of  Copula Link Based Survival Model(s) \citep{Marra2020} implemented in \texttt{GJRM} package. Before describing the algorithm, we will briefly introduce  the class of copula survival models  and  then discuss the variable selection algorithm and the implemented ranking measures in detail.
Readers interested in the technical details and analytical derivations of the algorithm are referred to \cite{Petti2024sub}. For further insights into the Copula Link Based Survival Model(s), a comprehensive  reference is \cite{Marra2020}.

\subsection{Copula Link Based Survival Model(s)}
\label{Ch:modelformulation}
Let $ \left(T_{1i}, T_{2i}\right) $  and $ \left(C_{1i}, C_{2i}\right) $ be a pair of survival times and censoring times, respectively, for unit $i$, with $i=1,\dots,n$, where $n$ is the sample size. Moreover, $ \left(C_{1i}, C_{2i}\right) $ are assumed to be independent of $ \left(T_{1i}, T_{2i}\right) $.

Let $\mathbf{x}_i$ be the $i^{\text{th}}$ row vector of the design matrix $\mathbf{X}\in \mathbb{R}^{n\times p}$,  where $p$ is the number of covariates. 


The conditional marginal survival functions for $T_{\nu i}$ and the conditional joint survival function are, respectively, given by $S_{\nu}(t_{\nu i}  \mid \mathbf{x}_{\nu i}; \boldsymbol{\beta}_{\nu})=P\left(T_{\nu i}>t_{\nu i} \mid \mathbf{x}_{\nu i}; \boldsymbol{\beta}_{\nu}\right)$ for $\nu=1,2$, and 

\begin{equation}
    S\left(t_{1 i}, t_{2 i} \mid \mathbf{x}_{i}; \boldsymbol{\delta}\right)=P\left(T_{1 i}>t_{1 i}, T_{2 i}>t_{2 i} \mid \mathbf{x}_{i}; \boldsymbol{\delta}\right).
\end{equation}

Then, the marginal survivals for the observed time values $t_{1i}$ and $t_{2i}$  are linked by a copula function $C(\cdot):[0,1]^2 \to [0,1]$:
\begin{equation}
\label{eq:jointSurvival}
S\left(t_{1 i}, t_{2 i} \mid \mathbf{x}_{i}; \boldsymbol{\delta}\right)=C\left(S_{1}\left(t_{1 i} \mid \mathbf{x}_{1 i}; \boldsymbol{\beta}_{1}\right), S_{2}\left(t_{2 i} \mid \mathbf{x}_{2 i}; \boldsymbol{\beta}_{2}\right); m\left\{\eta_{3 i}\left(\mathbf{x}_{3 i}; \boldsymbol{\beta}_{3}\right)\right\}\right),
\end{equation}
where the main elements of the Equation \eqref{eq:jointSurvival} are summarized in the Table \ref{tab:elementsofcopula}.

\begin{figure}[t!]
\centering
\includegraphics[height=0.6\textheight, keepaspectratio]{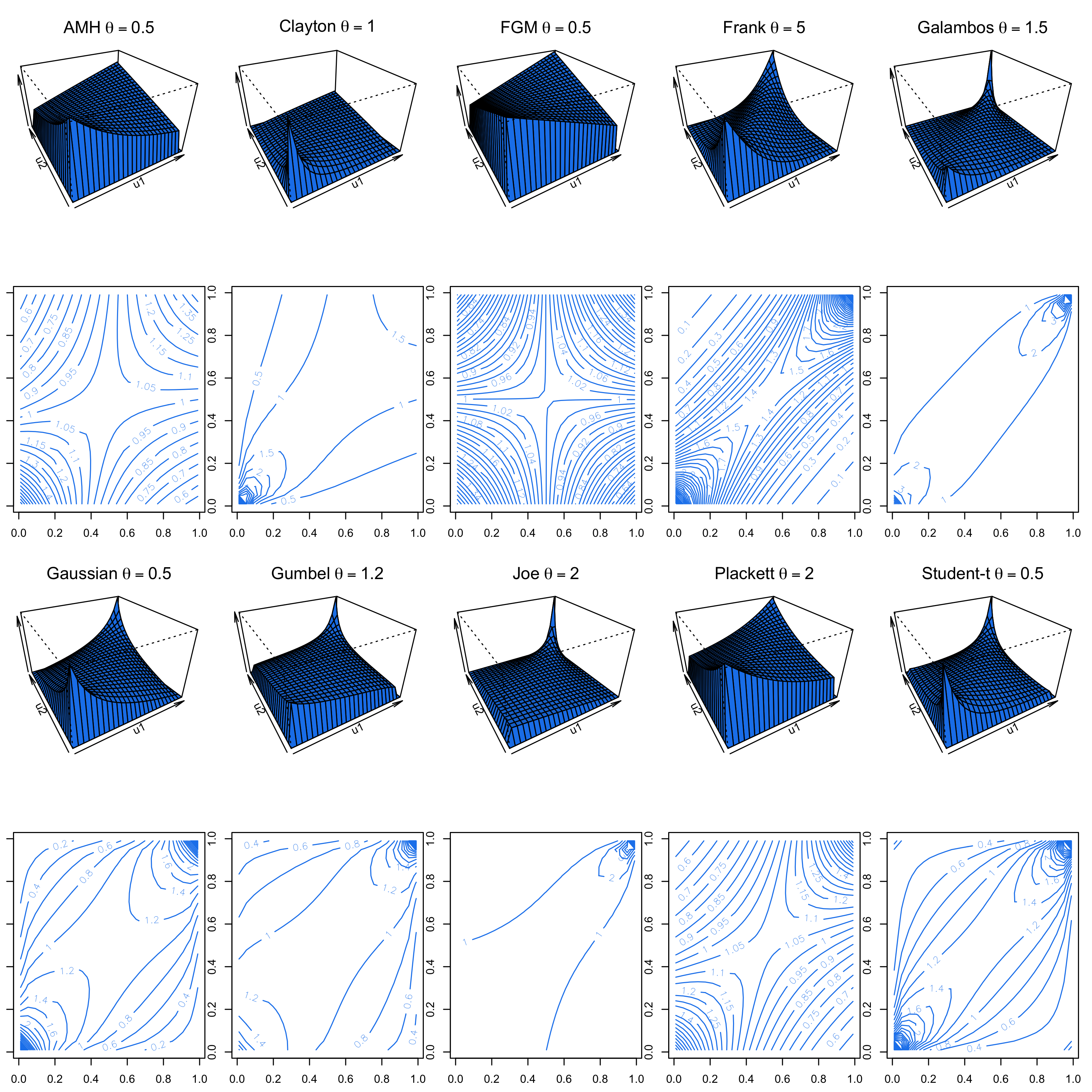}

 \caption{Bivariate copula PDFs presented in Table \ref{tab:copulafamilies}.
        Visualization of copulas in 3D and contour form. Each column represents a specific copula with its parameter $\theta$ indicated above. The first and third rows present the 3D density surfaces of the copulas, while the second and fourth rows show the corresponding contour plots. The copulas included are: AMH, Clayton, FGM, Frank, Galambos, Gaussian, Gumbel, Joe, Plackett, and Student-t, with parameters defining the dependency structure between variables. 
    }
\label{fig:copulaImage}
\end{figure}

The marginal survivals $S_1(\cdot)$ and $S_2(\cdot)$ are modeled by generalized survival or link-based models \citep{Liu2018, Royston2002}, leading to 

$$g_{\nu}\{S_{\nu}\left(t_{\nu i} \mid \mathbf{x}_{\nu i}; \boldsymbol{\beta}_{\nu})\right\}=\eta_{\nu i}\left(t_{\nu i}, \mathbf{x}_{\nu i}; \boldsymbol{\beta}_{\nu}\right),$$ 

where $g(\cdot):[0,1]\to \mathbb{R}$ is a  link function   and $\eta_{\nu i}\left(t_{\nu i}, \mathbf{x}_{\nu i}; \boldsymbol{\beta}_{\nu}\right) \in \mathbb{R}$, for $\nu=1,2$, are the additive predictors that  must include baseline functions of time (or a stratified set of functions of time) as clarified in \cite{Marra2020}. An overview of the choices for copulas and margins is presented in Figure \ref{fig:copulaImage} and Tables \ref{tab:copulafamilies} and \ref{tab:marl}. Thus, the three additive predictors can be written as:
\begin{equation}
\label{eq:AdditivePredictor}
\eta_{\nu i}=\beta_{\nu 0}+\sum_{k_{\nu}=1}^{K_{\nu}} s_{\nu k_{\nu}}\left(\mathbf{z}_{\nu k_{\nu} i}\right), \quad i=1, \ldots, n; \quad \nu=1,2,3,
\end{equation}

where $\beta_{\nu 0} \in \mathbb{R}$ denotes an overall intercept, $\mathbf{z}_{\nu k_{\nu} i}$ is the $k_{\nu}^{t h}$ sub-vector of the complete vector $\mathbf{z}_{\nu i}$ and the $K_{\nu}$ functions $s_{\nu k_{\nu}}\left(\mathbf{z}_{\nu k_{\nu}}\right.$) represent generic effects which are chosen according to the type of covariate(s), for $\nu=1,2$, $\mathbf{z}_{\nu k_{\nu} i}=(\mathbf{x}_{\nu k_{\nu} i}, t_{\nu k_{\nu} i})$ and for $\nu=3$, $\mathbf{z}_{3 k_{\nu} i}=\mathbf{x}_{3 k_{\nu} i}$. The main difference  between $\eta_\nu$ ($\nu=1,2$) and $\eta_3$ is that the first two must include baseline functions of time, and $t_{\nu_i}$ can be treated as regressor. We also note that the sets of covariates in the three margins may be the same but are not necessarily so.
For a more in-depth theoretical and methodological exploration as well as a comprehensive overview of the \texttt{GJRM} package, please refer to the following references:  \cite{Marra2020, Man:GJRM}.

\begin{table}[]
\renewcommand{\arraystretch}{1.3}
\begin{center}
{\centering
\begin{tabular}{p{3.5cm}p{10cm}}
\hline
Component & Description \\ \hline
& \\
$\boldsymbol{\delta}^{\top} \in \mathbb{R}^{W}$ & Vector of parameters containing $\boldsymbol{\beta}_1 \in \mathbb{R}^{W_1}$, $\boldsymbol{\beta}_2 \in \mathbb{R}^{W_2}$, $\boldsymbol{\beta}_3 \in \mathbb{R}^{W_3}$, with $W=\sum_{\nu=1}^3 W_{\nu}$. \\ 
$\mathbf{x}_i^{\top} \in \mathbb{R}^{W}$ & Vectors of covariates containing $\mathbf{x}_{1i} \in \mathbb{R}^{W_1}$, $\mathbf{x}_{2i} \in \mathbb{R}^{W_2}$, $\mathbf{x}_{3i} \in \mathbb{R}^{W_3}$, which can be sub-vectors of (or equal to) $\mathbf{x}_i$, with $W=\sum_{\nu=1}^3 W_{\nu}$.  \\ 
$C(\cdot):[0,1]^2\to[0,1]$ & The copula function captures the potentially varying conditional dependence of $\left(T_{1i}, T_{2i}\right)$ across observations. \\ 
$m(\cdot)$ & Inverse monotonic and differentiable link function ensuring that the dependence parameter lies in a proper range. \\ \\ \hline
\end{tabular}
}
\end{center}
\caption{Elements of the Bivariate Survival Copula Link-based Additive model expressed in Equation \eqref{eq:jointSurvival}.}
\label{tab:elementsofcopula}
\end{table}

\begin{table}[t]
\begin{center}
\renewcommand{\arraystretch}{2.5}
{\footnotesize 
\begin{tabular}{lccc}
\hline
Copula & $C(u_1,u_2;\theta)$ & Range of $\theta$ \\ \hline

AMH (\texttt{"AMH"}) & $\frac{u_1u_2}{1-\theta(1-u_1)(1-u_2)}$ & $[-1,1]$ \\ 

Clayton (\texttt{"C0"}) & $\left( u_1^{-\theta} + u_2^{-\theta} -1 \right)^{-1/\theta} $ & $(0,\infty)$ \\ 

FGM (\texttt{"FGM"}) & $u_1u_2\left\{1 + \theta(1 - u_1)(1 - u_2)\right\}$ & $[-1,1]$ \\ 

Frank (\texttt{"F"}) & $-\theta^{-1} \log \left\{ 1+(\exp\left\{-\theta u_1\right\}-1)\right.$ \\
& $\left.(\exp\left\{-\theta u_2\right\}-1)/(\exp\left\{-\theta\right\}-1) \right\} $ & $\mathbb{R}\backslash \left\{0\right\}$ \\ 

Galambos (\texttt{"GAL"}) & $u_1u_2\exp\left[\left\{(-\log u_1)^{-\theta} \right.\right.$ \\ 
& $\left.\left. + (-\log u_2)^{-\theta} \right\}^{-1/\theta} \right]$  & $(0,\infty)$ \\ 

Gaussian (\texttt{"N"}) & $\Phi_2\left(\Phi^{-1}(u_1),\Phi^{-1}(u_2);\theta \right) $ & $[-1,1]$ \\ 

Gumbel (\texttt{"G0"}) & $\exp\left[-\left\{(-\log u_1)^\theta \right.\right.$ \\ 
& $\left.\left. + (-\log u_2)^\theta \right\}^{1/\theta} \right]$  & $[1,\infty)$ \\ 

Joe (\texttt{"J0"}) & $1-\left\{ (1-u_1)^\theta+ (1-u_2)^\theta\right.$ \\ 
& $\left. - (1-u_1)^\theta(1-u_2)^\theta \right\}^{1/\theta}$ & $(1,\infty)$ \\ 

Plackett (\texttt{"PL"}) & $\left(Q-\sqrt{R}\right)/\left\{2(\theta-1)\right\}$ & $(0,\infty)$ \\ 

Student's t (\texttt{"T"}) & $t_{2,\zeta}\left(t_{\zeta}^{-1}(u_1),t_{\zeta}^{-1}(u_2);\zeta,\theta \right) $ & $[-1,1]$ \\ 
\hline
\end{tabular}
}
\end{center}
\caption{Description of copulae available in \texttt{BRBVS} package, with corresponding parameter range of association parameter $\protect\theta$. $\Phi_2(\cdot,\cdot;\protect\theta)$ denotes the cumulative distribution function (cdf) of the standard bivariate normal distribution with correlation coefficient $\protect\theta$, and $\Phi(\cdot)$ the cdf of the univariate standard normal distribution. $t_{2,\protect\zeta}(\cdot,\cdot;\protect\zeta,\protect\theta)$ indicates the cdf of the standard bivariate Student-t distribution with correlation $\protect\theta$ and fixed $\protect\zeta\in(2,\infty)$ degrees of freedom, and $t_{\protect\zeta}(\cdot)$ denotes the cdf of the univariate Student-t distribution with $\protect\zeta$ degrees of freedom.}
\label{tab:copulafamilies}
\end{table}


\begin{table}[htbp]
\footnotesize
\begin{center}
\renewcommand{\arraystretch}{2.5}
\begin{tabular}{lcccc}
\hline
Model & Link $g(S)$ & Inverse link $g^{-1}(\eta)=G(\eta)$ & $G'(\eta)$ \\ \hline
Prop.hazards (\texttt{"PH"}) & $\log \left\{-\log(S) \right\}$  & $\exp\left\{-\exp(\eta)\right\}$ &  $-G(\eta)\exp(\eta)$ \\
Prop.odds (\texttt{"PO"}) & $-\log\left(\frac{S}{1-S}\right)$ & $\frac{\exp(-\eta)}{1+\exp(-\eta)}$ & $-G^2(\eta)\exp(-\eta)$  \\
probit (\texttt{"probit"}) & $-\Phi^{-1}(S)$ & $\Phi(-\eta)$ & $-\phi(-\eta)$\\
\hline 
\end{tabular}
\end{center}
\caption{Description of link functions available in \texttt{BRBVS} package. $\Phi$ and $\phi$ are the cumulative distribution and density functions of a univariate standard normal distribution.}
\label{tab:marl}
\end{table}

\clearpage
\subsection{The Bivariate Ranking Based Variable Selection Algorithm }
 \label{sec:BRBVS}

The BRBVS algorithm implemented in the package is here  formally described, also  underling the main differences between the BRBVS and the RBVS algorithm by \cite{Baranowski2020}. The differences are mainly two: a) the BRBVS extends the RBVS to the copula bivariate and survival domains; b) it jointly accounts for the dimension reduction of the dataset, considering the dependence between the two survival margins. This last point has been faced by introducing a new ranking measure for the covariates, as largely presented in \cite{Petti2024sub}. In the following, the two main steps of the algorithm are illustrated: the variable ranking and the variable selection.

\vspace{5pt}
\textbf{ Step 1. Variable ranking}\\
Let $\mathcal A_\nu\subset (1,2,\ldots, p)$, for $\nu=1,2$, be the indices that identify a subset of covariates contained in the design matrix $\bf X$ and included in the first and second margin,  and let $|\mathcal A_\nu|=k_{\nu}$ be their cardinality, respectively. Furthermore, let ${\bf z}=({\bf z}_1, \ldots, {\bf z}_n)$ be a matrix where ${\bf z}_i=(t_{i1}, t_{i2}, x_{i1}, \ldots, x_{ip})$ for $i=1,2,\ldots, n$, and let $  \mathcal R_{\nu j}({\bf z})$ be the ranking assigned to the covariate $X_j$, for $j=1,2,\ldots, p$, in $\eta_\nu$, based on an estimated measure $\hat\omega_{\nu j}({\bf z})$. This measure is jointly defined for both $\eta_1$ and $\eta_2$, to assess the importance of the covariates in each margin and is defined such that $\hat\omega_{\nu j}>\hat\omega_{\nu (j+1)}$, for $j=1,\ldots, (p-1)$ and $\nu=1,2$. 
More details about the choice of this measure are given the the following.

Under the theoretical framework in Section \ref{Ch:modelformulation}, we can define, for each $\eta_\nu$, a unique top-ranked set of covariates such that the probability that the top-$k_\nu$ variables are included in $\eta_\nu$ is:
\begin{equation}\label{eq:pi_v}
\pi_\nu(\mathcal A_\nu)=P\bigg(\left\{\mathcal R_{\nu_1}({\bf z}), \mathcal R_{\nu_2}({\bf z}),\ldots, \mathcal R_{\nu_{k_\nu}}({\bf z})\right\}=\mathcal A_\nu\bigg),
\end{equation}
with $\pi_\nu(\mathcal A_\nu)=1$, if $\mathcal A_\nu=\emptyset$.

To estimate the probability $\pi_\nu(\mathcal A_\nu)$, in line with \cite{Baranowski2020},
we consider $B$ bootstrap replicates and for each of them we draw uniformly without replacement $r=\lfloor n/m \rfloor$ random samples of size $m$  from $\bf{X}$ (with $\lfloor n/m \rfloor$ the integer part of the ratio). Then, the estimate of the probability $\pi_\nu(\mathcal A_\nu)$ for the sets of covariates $\mathcal A_\nu$ is given by:

\begin{equation}\label{eq:hat_pi_v}
\hat\pi_{\nu,m}(\mathcal A_\nu)=B^{-1}\sum_{b=1}^Br^{-1}\sum_{q=1}^r{\bf 1}(\mathcal A_\nu|I_{b_q}),
\end{equation}
where ${\bf 1}(\cdot)$ is an indicator function which assumes value 1 when the covariates indexed in $\mathcal A_\nu$ are top-ranked and 0 otherwise; $I_{b_q}$ is the $q^{\text{th}}$ subsample extracted from the data, for $q=1,\ldots, r$, $\mathcal A_\nu|I_{b_j}$ is the subset of $k_\nu$ covariates of $\eta_\nu$, with $k_\nu=0,1,\ldots, p-1$, whose ranking is computed through the subsample $I_{b_q}$.

These probabilities allow us to define the top-ranked covariates for the two margins $\eta_\nu$:
\begin{equation}\label{eq:max_pi}
\hat{\mathcal A}_{\nu,m,k_v}=\underset{\mathcal A_\nu\in \Omega_{k_\nu}}{\argmax}\quad\hat\pi_{\nu,m}(\mathcal A_\nu),
\end{equation}
with $\Omega_{k_\nu}$ the set of all permutations of $\{1,\ldots, k_\nu\}$.

In practice, the variables included in \(\hat{\mathcal{A}}_{\nu,m,k_\nu}\) are those with the highest probability, and correspond to the covariates that most frequently are at the top of the ranking in the \(B \cdot r\) sub-samples. 

Shortly, in this first step $(2\times B)$ rankings of the covariates are obtained from the bootstrap replicates, (that, in other words,  correspond to $B$ rankings obtained for each of the two margins); they are used to estimate the probability \eqref{eq:hat_pi_v}, and then from the maximization \eqref{eq:max_pi} the top-ranked covariates (one for each margin) are defined.

\vspace{5pt}
\textbf{ Step 2. Variable selection}\\
From Step 1. two rankings of cardinality $p$ are obtained (one for each margin).  The variable selection may then be  performed, for example, by fixing a threshold  to  $k_\nu$ such that all ranked covariates exceeding this threshold are discarded. Alternatively, the threshold may be fixed to the estimated probability $\hat\pi_{\nu,m}(\mathcal A_\nu)$, for $\nu=1,2$, and then the set of covariates that have an estimated probability less than the given threshold are discarded. In practice, it implies that the selected covariates are the first $k_\nu$ in the ranking or, alternatively, those with probability $\hat\pi_{\nu,m}(\mathcal A_\nu)$ greater than a given threshold value.
Unfortunately, in the presence of the model \eqref{eq:jointSurvival} the definition of a threshold has additional difficulties due to the presence of two subsets of covariates to select and  the application of the same threshold value may not be an appropriate choice in this case. In practice, the order of magnitude of the probabilities \eqref{eq:hat_pi_v} may be quite different for the two margins, and then the definition of a threshold can lead to over(under)-selecting the variables. By contrast, even the definition of  pre-specified thresholds can lead to a subjective variable selection.

For these reasons, following the idea in \cite{Baranowski2020}, for each $\eta_\nu$ we select the subset $\hat s_\nu$ of important variables such that:

\begin{equation}\label{eq:hat_s_v}
\hat{s}_\nu=\underset{k_\nu=0, \ldots, k_{\max}-1}{\argmin}\frac{[\hat\pi_{\nu,m}(\hat{\mathcal A}_{\nu, m, k_\nu+1})]^\tau}{\hat\pi_{\nu,m}(\hat{\mathcal A}_{\nu, m, k_\nu})},
\end{equation}
with $\tau\in(0,1]$ and $k_{\max}$ the maximum number of covariates considered to compute the ratio \eqref{eq:hat_s_v}, for $\nu=1,2$.

In practice, for each $\eta_\nu$ and given $\tau$, the number of selected variables is chosen by looking at the ratio in \eqref{eq:hat_s_v} and then $k_\nu$ variables are included in $\hat{s}_\nu$ such that $[\hat\pi_{\nu,m}(\hat{\mathcal A}_{\nu, m, k_\nu+1})]^\tau/\hat\pi_{\nu,m}(\hat{\mathcal A}_{\nu, m, k_\nu})$ drastically decreases concerning $[\hat\pi_{\nu,m}(\hat{\mathcal A}_{\nu, m, k_\nu})]^\tau/\hat\pi_{\nu,m}(\hat{\mathcal A}_{\nu, m, k_\nu-1})$.

The ratio \eqref{eq:hat_s_v} allows us to select the relevant covariates that also correspond to the subset of variables included in $\bf X$ with higher ranking. Furthermore, note that \(\hat{s}_\nu\) is estimated for each of the two margins and then their cardinality does not necessarily have to be equal.

\begin{sidewaysfigure}
\centering
\includegraphics[height=0.60\textheight, keepaspectratio]{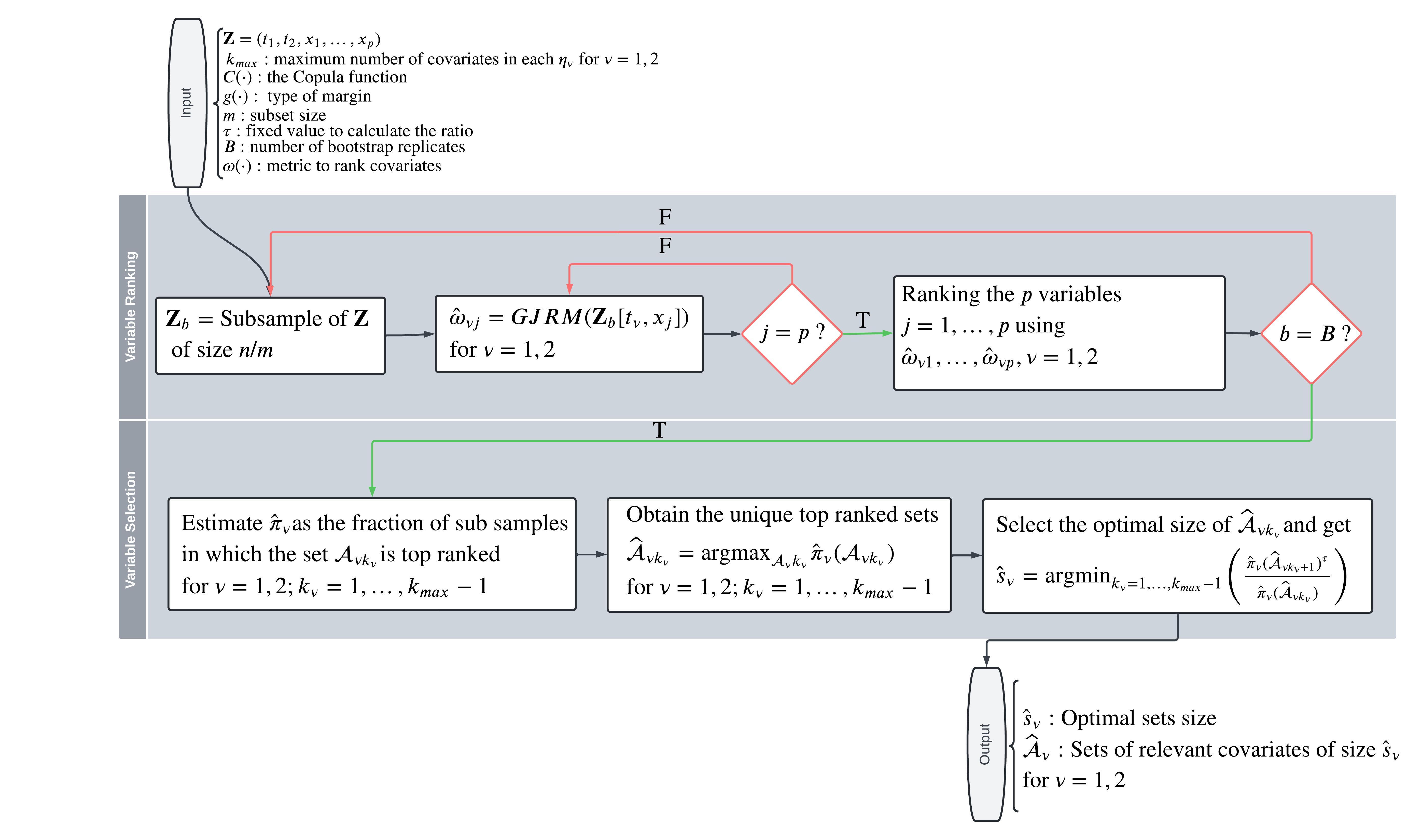}
\caption{Flow chart representation of the BRBVS algorithm presented in Algorithm \ref{alg:BRBVS}.}
\label{fig:BRBVS flowchart}
\end{sidewaysfigure}

The structure of the BRBVS algorithm in described in Algorithm \ref{alg:BRBVS} and can be sketched in four steps (Figure \ref{fig:BRBVS flowchart}): the first two for the variable ranking, and the last two for the variable selection.

\begin{algorithm}
 \caption{BRBVS Algorithm.}\label{alg:BRBVS}
 \SetAlgoLined
 {{\bf Input:} $BRBVS(\boldsymbol{Z}, k_{max}, C(\cdot), g(\cdot),  m, \tau, B, \omega(\cdot))$}
 \BlankLine
 \# $\boldsymbol{Z}$: $[n\times (p+2)]$ data matrix with $\mathbf{z}_i=(t_{i1}, t_{i2}, x_{i1}, \ldots, x_{ip})$;\\
  \# $k_{max}$: maximum number of covariates in each $\eta_\nu$, for $\nu=1,2$;\\
  \# $C(\cdot)$: the Copula function (selected among them implemented in the \href{https://cran.r-project.org/package=GJRM}{\texttt{GJRM} package});\\
  \# $g(\cdot)$: type of margin (selected among them implemented in the \href{https://cran.r-project.org/package=GJRM}{\texttt{GJRM} package})\\
 \# $m$: subset size;\\
 \# $\tau$: fixed value to calculate the ratio \\
 \# $B$: number of bootstrap replicates;\\
 \# $\omega(\cdot)$: metric to rank the covariates;\\
 \BlankLine
 \For{$b\leftarrow 1$ \KwTo $B$}{
  {\bf Step 1.1:} Draw $r$ subsamples from $\boldsymbol{Z}$ uniformly without replacement, where $r=\lfloor n/m \rfloor$;\\
  {\bf Step 1.2:} Estimate $\hat{\omega}_{\nu j}(\boldsymbol{Z})$, for $j = 1, \ldots, p$ and $\nu = 1, 2$, by using the selected metric $\omega(\cdot)$ and the output of the  \href{https://cran.r-project.org/package=GJRM}{\texttt{GJRM} package};\\
  {\bf Step 1.3:}  in each margin rank the corresponding variables such that  $\hat\omega_{\nu j}>\hat\omega_{\nu (j+1)}$, for $j=1,2,\ldots, p$;\\
 }
 {\bf Step 1.4:} Estimate the probabilities $\hat{\pi}_{\nu, m}$ $(\mathcal{A}_{\nu})$ as in \eqref{eq:hat_pi_v}, for $\nu=1,2$;\\
 {\bf Step 2:} Given $\tau$, select the sets of important covariates $\hat{s}_\nu$, for $\nu=1,2$, such that the ratio \eqref{eq:hat_s_v} is minimized, also defining the corresponding $\hat{\mathcal{A}}_{1}, \hat{\mathcal{A}}_{2}$;\\
 \BlankLine
 {{\bf Output:} $(\hat{s}_1, \hat{s}_2, \hat{\mathcal{A}}_{1}, \hat{\mathcal{A}}_{2})$}
\end{algorithm}

The algorithm presented takes as inputs: 1) the dataset; 2) $ k_{\text{max}} $, as clarified in Equation \eqref{eq:hat_s_v}, represents the maximum number  of variables selected in each margin; 3) the Copula function $C(\cdot)$, as in equation \eqref{eq:jointSurvival}; 4) the type of margin $g(\cdot)$; 5) the value $ m $, which is the dimension of the subsamples used in the ranking step and usually fixed to  half of the sample size;  6) \( \tau \), which is set by default to 0.5 and guides the decay of the threshold in Equation \eqref{eq:hat_s_v}; 7) the number of bootstrap replicates (by default fixed to $ B = 50 $ because the empirical evidence suggests this value as optimal); 8) metric to be used to compute $\hat\omega_{\nu j}$, for $\nu=1,2$ and $j=1,2,\ldots, p$.

\clearpage
 \subsubsection{Ranking measure}
One of the most crucial elements in the proposed algorithm is the definition of an appropriate measure $\omega$ to rank the covariates, which needs numerically summarize the importance of the  covariates in fitting the two margins $\nu=1,2$. The \texttt{BRBVS} implements  three measures presented  in Table \ref{tab:measures}. However, as  clarified in the following, it is recommended to use only the first two, namely \texttt{FIM} and \texttt{Abs}. The first measure, \texttt{FIM}, originally proposed in \cite{Petti2024sub},  is based of the use of the Fisher Information Matrix of the class of models discussed in Section \ref{sec:Framework}, while the second measure, \texttt{Abs},   establishes the variable ranking on the absolute value of the estimated  coefficients.

In detail, the \texttt{FIM} makes use of the Fisher Information to isolate the contribution of the $j^{\text{th}}$ covariate to the $\nu^{\text{th}}$ survival function by accounting for the sharpness of the likelihood in the direction of the parametric effect. A high value of $\mathbb{E}\bigg[ \frac{\partial^2 \ell(\boldsymbol \delta)}{\partial \beta_{\nu j} \partial \beta_{\nu j}} \bigg]$ denotes a sharp curvature of the likelihood, indicating greater accuracy in the estimate \citep{amari2012differential}. 
This measure has proven to be the best among the three in Table \ref{tab:measures}, as will be clarified through a simulation study in Section \ref{sec:simulation}.

The absolute value of the coefficients \texttt{"Abs"} is a measure that was originally considered in \cite{Baranowski2020} for the class of linear models and for the univariate version of the algorithm presented in this section.

The copula entropy \texttt{"CE"} was recently proposed in univariate contexts in \cite{cheng2022variable, frenay2013mutual}, and in survival contexts in \cite{ma2022copula}. Unfortunately, this measure has proven to be inadequate for variable ranking in bivariate domain, and more details in this regard are presented in the Supplementary Material of \cite{Petti2024sub}.

\begin{table}[]
\renewcommand{\arraystretch}{1.3}
\begin{center}
{\centering
\begin{tabular}{p{7cm}p{7cm}}
\hline
Name  & Formula \\ \hline
Fisher Information Measure ("\texttt{FIM}")& ${\beta}^2_{\nu j}  \mathbb{E}\bigg[ \frac{\partial^2  \ell(\boldsymbol \delta) }{\partial \beta_{\nu j} \partial \beta_{\nu j} } \bigg]$\\
Absolute Value of the Coefficients (``\texttt{Abs}'')& $|{\beta}_{\nu j}|$ \text{where} $|\cdot|=$ \text{Absolute value}\\
Copula Entropy ("\texttt{CE}")&\( MI_{\nu j}(T_\nu, X_j)=\int_{[0,1]^2} C(\mathbf{u})\log{C(\mathbf{u})} d \mathbf{u} \), where $ \mathbf{u}=\big[F_{\eta_\nu}(z), $ $ F_{X_j}(x)\big] $ \\ \hline
\end{tabular}
}
\end{center}
\caption{Measures $\omega(\cdot)$ implemented in the \texttt{BRBVS} package. Here, $\beta_{\nu j}$ represents the linear effect of the covariate \( j \) for the margin \(\nu = 1,2\). The term $\mathbb{E}\bigg[ \partial^2 \ell(\boldsymbol \delta) /\partial \beta_{\nu j} \partial \beta_{\nu j} \bigg]$ is the $(j,j)^{\text{th}}$ diagonal element in position \([j,j]\) of the Fisher information matrix; \(MI_{\nu j}(T_\nu, X_j)\) denotes the  Mutual Information (Copula Entropy) between the \(\nu\)-th time to event and the \(j\)-th covariate.}
\label{tab:measures}
\end{table}

 \clearpage
\section{The BRBVS package}
\label{sec:packageDetails}
The \texttt{BRBVS} package includes a set of functions related to the Bivariate Variable Ranking Based Variable Selection (BRBVS) method introduced in \cite{Petti2024sub}. These functions are useful for performing the variable selection in the presence of the bivariate censored survival data and obtaining both numerical and visual outputs. In addition to the bivariate algorithm discussed in Section \ref{sec:Framework}, the package implements backward and forward selection methods, as well as a function for selecting the best link function.  In the following section, we will describe the functions implemented in this package.

\subsection{Package overview}
The package is available on the Comprehensive
R Archive Network (CRAN) at \href{https://cran.r-project.org/web/packages/BRBVS/index.html}{https://cran.r-project.org/web/packages/BRBVS} and can be installed and loaded in the usual manner: 

\begin{verbatim}
R> install.packages("BRBVS")
R> library(BRBVS)
\end{verbatim}

To illustrate the  package, we use the AREDS dataset, which is a sample of real-world bivariate interval-censored data consisting of 629 subjects. This dataset includes four non-genetic covariates (\texttt{SevScale1E, SevScale2E, ENROLLAGE}) and one genetic covariate (\texttt{rs2284665}). The data are derived from the Age-Related Eye Disease Study (AREDS) \citep{AREDSstudy1999}.

First, we separate the times to event variables and censoring indicators from the covariates, storing the former in \texttt{Y} and the latter in \texttt{X}. Next, we standardize the numeric covariates to project them into a comparable space, ensuring that covariates with different scales contribute equally to the analysis. Additionally, we encode  \texttt{rs2284665} into three distinct levels: \texttt{GG}, \texttt{GT}, and \texttt{TT}, using dummy variables for each category.

\begin{verbatim}
R> data(AREDS)

R> dim(AREDS)
[1] 628  11


R> Y <- AREDS[, c('t11', 't12', 't21', 't22', 'cens1', 'cens2', 'cens')]
R> head(Y)

       t11 t12     t21 t22 cens1 cens2 cens
1   0.0001 2.0  0.0001 2.0     I     I   II
3   0.0001 2.0  5.9000 9.3     I     I   II
5   8.0000 9.1 10.0000  NA     I     R   IR
7   3.0000 4.1  3.0000 4.1     I     I   II
9   4.8000 5.8  0.0001 1.8     I     I   II
11 10.0000  NA 10.0000  NA     R     R   RR


R> X <-AREDS[, c('SevScale1E', 'SevScale2E', 'ENROLLAGE', 'rs2284665')]
R> X$SevScale1E <- scale(as.numeric(X$SevScale1E))
R> X$SevScale2E <- scale(as.numeric(X$SevScale2E))  
R> X$ENROLLAGE <-  scale(X$ENROLLAGE)

R> X$GG <- ifelse(X$rs2284665 == 0, 1, 0)  # Genotype GG
R> X$GT <- ifelse(X$rs2284665 == 1, 1, 0)  # Genotype GT
R> X$TT <- ifelse(X$rs2284665 == 2, 1, 0)  # Genotype TT

R> head(X)

   SevScale1E SevScale2E  ENROLLAGE rs2284665 GG GT TT
1   0.1465588  1.7167427 -0.4865267         1  0  1  0
3   0.9332163 -1.4287407 -0.2956723         0  1  0  0
5   0.9332163  0.9303719 -0.8873208         0  1  0  0
7   0.1465588  0.1440010 -1.1354315         1  0  1  0
9   0.9332163  0.9303719 -0.3529286         1  0  1  0
11  0.1465588  0.1440010  0.7921975         0  1  0  0

\end{verbatim}

Before running the BRBVS algorithm using the \texttt{BRBVS::BRBVS()} function, it is important to identify the link functions and copula functions that best fit the data. 

In the \texttt{BRBVS} package, the \texttt{BRBVS::Select\_link\_BivCop()} function identifies the best link functions for the first and second time-to-event variables, as outlined in Table~\ref{tab:marl}. The key parameters are:

\begin{itemize}
    \item \texttt{data}: A data frame containing the dataset (e.g., AREDS data).
    \item \texttt{cens1} and \texttt{cens2}: Censoring indicators for the first and second time-to-event variables, respectively.
    \item \texttt{lowerBt1}, \texttt{lowerBt2}, \texttt{upperBt1}, and \texttt{upperBt2}: Character strings specifying the names of the lower and upper bounds for the first and second time-to-event variables.
    \item \texttt{measure}: Criterion to minimize during selection, either \texttt{AIC} (default) or \texttt{BIC}.
    \item \texttt{eta1} and \texttt{eta2}: Formulas for the survival models of the first and second time-to-event variables, defaulting to \texttt{NULL}.
    \item \texttt{input\_equation}: Logical. If \texttt{TRUE}, uses the provided \texttt{eta1} and \texttt{eta2} formulas. If \texttt{FALSE} (default), generates formulas using all predictors in \texttt{data}.
\end{itemize}

Further details about the arguments are provided in Table~\ref{tab:infoSelectLink}, and the algorithm is summarized in Algorithm~\ref{alg:select_link_BivCop}.

\begin{table}[ht]
\renewcommand{\arraystretch}{1.3}
\begin{center}
{\footnotesize
\begin{tabular}{l p{10cm}}
\hline
{Arguments} & {Description} \\
\hline
\texttt{data} & A data frame containing the times to event \texttt{t11}, \texttt{t12}, \texttt{t21}, \texttt{t22}, censoring variables \texttt{cens1}, \texttt{cens2}, and covariates. \\
\texttt{cens1} & Character. Censoring indicator for the first time to event. \\
\texttt{cens2} & Character. Censoring indicator for the second time to event. \\
\texttt{lowerBt1} & Character. Name of the lower bound for the first time to event. \\
\texttt{lowerBt2} & Character. Name of the lower bound for the second time to event. \\
\texttt{upperBt1} & Character. Name of the upper bound for the first time to event. \\
\texttt{upperBt2} & Character. Name of the upper bound for the second time to event. \\
\texttt{measure} & Character. Measure to be minimized during the selection process. Either the Akaike information criterion \texttt{("AIC")} or Bayesian information criterion \texttt{"BIC"}. Default is \texttt{"AIC"}. \\
\texttt{eta1} & Formula for the first survival model equation. Default is \texttt{NULL}. \\
\texttt{eta2} & Formula for the second survival model equation. Default is \texttt{NULL}. \\
\texttt{input\_equation} & Logical. If \texttt{TRUE}, uses the provided \texttt{eta1} and \texttt{eta2} formulas. If \texttt{FALSE}, generates formulas using all predictors in data. Default is \texttt{FALSE}. \\
\hline
\end{tabular}
}
\end{center}
\caption{\texttt{BRBVS::Select\_link\_BivCop()} argument description.}
\label{tab:infoSelectLink}
\end{table}

\begin{algorithm}
{\footnotesize
 \caption{Select the Best Link  function algorithm  implemented in \texttt{BRBVS}.}\label{alg:select_link_BivCop}
 \SetAlgoLined
 {{\bf Input:} $SelectLink(\mathbf{Z}$, $IC$})
 \BlankLine
 \# $\mathbf{Z}$: $[n \times (p+2)]$ data matrix with $\mathbf{z}_i=(t_{1i}, t_{2i}, x_{i1}, \dots, x_{ip})$;\\
 \# $IC$: Information Criterion for model selection, either AIC or BIC;\\
 \BlankLine
 {{\bf Output:} Optimal link functions for $\eta_1$ and $\eta_2$}
 \BlankLine
 \# Initialize best $IC$ values and corresponding link functions for $\eta_1$ and $\eta_2$;\\
 $IC_{\text{best}, \eta_1} \gets \infty$; $IC_{\text{best}, \eta_2} \gets \infty$;\\
 $\text{link}_{\text{best}, \eta_1} \gets \emptyset$; $\text{link}_{\text{best}, \eta_2} \gets \emptyset$;\\
 \BlankLine
 \For{each link function $l \in \{$'PH', 'PO', 'Probit'$\}$}{
    \BlankLine
    \# Fit model $\eta_1$ using current link $l$ and calculate $IC$;\\
    $M_{\eta_1, l} \gets \text{GJRM}(\eta_1, \mathbf{Z}, l)$;\\
    $IC_{\eta_1, l} \gets IC(M_{\eta_1, l})$;\\
    \BlankLine
    \# Fit model $\eta_2$ using current link $l$ and calculate $IC$;\\
    $M_{\eta_2, l} \gets \text{GJRM}(\eta_2, \mathbf{Z}, l)$;\\
    $IC_{\eta_2, l} \gets IC(M_{\eta_2, l})$;\\
    \BlankLine
    \# Update best $IC$ and link function for $\eta_1$ if current $IC$ is better;\\
    \If{$IC_{\eta_1, l} < IC_{\text{best}, \eta_1}$}{
        $IC_{\text{best}, \eta_1} \gets IC_{\eta_1, l}$;\\
        $\text{link}_{\text{best}, \eta_1} \gets l$;\\
    }
    \BlankLine
    \# Update best $IC$ and link function for $\eta_2$ if current $IC$ is better;\\
    \If{$IC_{\eta_2, l} < IC_{\text{best}, \eta_2}$}{
        $IC_{\text{best}, \eta_2} \gets IC_{\eta_2, l}$;\\
        $\text{link}_{\text{best}, \eta_2} \gets l$;\\
    }
 }
 \BlankLine
 \textbf{return} $(\text{link}_{\text{best}, \eta_1},  \text{link}_{\text{best}, \eta_2})$;\\
 }
\end{algorithm}

\clearpage
\begin{verbatim}
R> Best_links <- BRBVS::Select_link_BivCop(data = AREDS, 
                   cens1, cens2,
                   lowerBt1 = "t11", lowerBt2 = "t21",
                   upperBt1 = "t12", upperBt2 = "t22",
                   measure = "AIC", 
                   eta1 = NULL, eta2 = NULL,
                   input_equation = FALSE)    
\end{verbatim}

 Part of the output of the \texttt{BRBVS::Select\_link\_BivCop()} function is shown below

\begin{verbatim}
    

R> print(Best_links)

Summary of Best Margins for Survival Analysis:
-------------------------------------------------
Survival 1:
Best Link Function: PO
AIC Value: 2163.371328
-------------------------------------------------
Survival 2:
Best Link Function: PO
AIC Value: 2272.726336
-------------------------------------------------    
\end{verbatim}

The output indicates that the best link function based on the AIC (Akaike information criterion) is the proportional odds \texttt{PO}.

At this stage, we can run the \texttt{BRBVS::BRBVS()} function. The key arguments are as follows:

\begin{itemize}
    \item \textbf{\texttt{y = Y}}: Specifies the response variables, including \texttt{t11}, \texttt{t12}, \texttt{t21}, \texttt{t22}, and the censoring indicators \texttt{cens1} and \texttt{cens2}.
    \item \textbf{\texttt{x = X}}: Specifies the covariates.
    \item \textbf{\texttt{kmax = 5}}: Sets the maximum number of variables to be selected, \(k_{\max}\). We recommend choosing \(k_{\max} < \min(n, p)\) to reduce computational burden.
    \item \textbf{\texttt{copula = "PL"}}: Specifies the copula to be used, in this case, a Plackett copula.
    \item \textbf{\texttt{margins = c("PO", "PO")}}: Indicates that both margins follow the proportional odds model.
    \item \textbf{\texttt{m = n/2}}: Sets the subsample size, where \texttt{n =629} is the sample size.
    \item \textbf{\texttt{tau = 0.5}}: Specifies the threshold for variable inclusion, as defined in Equation~\eqref{eq:hat_s_v}.
    \item \textbf{\texttt{n.rep = 50}}: Sets the number of bootstrap replications, \(B\).
    \item \textbf{\texttt{metric = 'FIM'}}: Specifies the measure \(\omega\), in this case, the Fisher Information Measure, as defined in Table~\ref{tab:measures}.
\end{itemize}

For more details, the arguments of the \texttt{BRBVS::BRBVS()} function are listed and explained in Table~\ref{tab:infoBRBVS}.

\begin{verbatim}
R> Bivrbvs<- BRBVS::BRBVS(y=Y, x=X, kmax=5, copula="PL", 
                    margins=c("PO","PO"),  m= 629/2, tau=0.5,
                    n.rep=50,  metric='FIM')
\end{verbatim}

\begin{table}[ht]
\renewcommand{\arraystretch}{1.3}
\begin{center}
{\footnotesize
\begin{tabular}{l p{10cm}}
\hline
{Argument} & {Description} \\
\hline
\texttt{y} & times to event and censoring matrix as a data frame. \\

\texttt{x} & Covariates matrix as a data frame. Input matrix containing the  variables. \\

\texttt{kmax} & Numeric. The maximum number of variables to be selected. Must be positive, non-zero, and less than or equal to the number of columns in \texttt{x}. \\

\texttt{copula} & Character. Type of copula employed in the algorithm. Must be one of the following types: bivariate normal (\texttt{"N"}), Clayton (\texttt{"C0"}), Galambos (\texttt{"GAL0"}), Joe (\texttt{"J0"}), Gumbel (\texttt{"G0"}), Frank (\texttt{"F"}), Ali-Mikhail-Haq (\texttt{"AMH"}), Farlie-Gumbel-Morgenstern (\texttt{"FGM"}), Student-t with degrees of freedom (\texttt{"T"}), Plackett (\texttt{"PL"}), and Hougaard (\texttt{"HO"}). Default is \texttt{"C0"}. Each copula can also be combined with a rotated version of the same family, allowing for modeling of negative and positive tail dependencies.\\

\texttt{margins} & Character. Type of margin employed in the algorithm. Must be one of \texttt{PH}, \texttt{PH}, \texttt{probit}. Default is \texttt{c(PH, PO)}. \\

\texttt{m} & Numeric. Subsample size, typically set to $n/2$ where $n$ is the number of observations. \\

\texttt{tau} & Numeric. A user-defined threshold for variable selection. Must be in the interval $(0,1)$, exclusive. Usually set equal to $0.5$  \citep[see][]{Baranowski2020} \\

\texttt{n.rep} & Integer. Number of Bootstrap replicates. Must be positive. \\

\texttt{metric} & Character. Specifies the metric used for ranking the variables. Must be one of Copula Entropy (\texttt{"CE"}), Fisher information measure (\texttt{"FIM"}), Absolute value of the coefficients (\texttt{"Abs"}). Default is \texttt{"FIM"}. \\
\hline
\end{tabular}
} 
\end{center}
\caption{ \texttt{BRBVS()} argument description.}
\label{tab:infoBRBVS}
\end{table}

\begin{table}[ht]
\renewcommand{\arraystretch}{1.3}
\begin{center}
{\footnotesize
\centering
\begin{tabular}{l p{10cm}}
\hline
{Function} & {Description} \\ \hline
\texttt{summary()} & Returns the hyperparameters and two sets of relevant variables with associated frequency of selection. \\ 
\texttt{plotBRBVS()} & Histogram of selected features against the relative frequency of selection. \\ \hline
\end{tabular}
}
\end{center}
\caption{Methods of \texttt{BRBVS} objects.}
\label{tab:methodsBRBVS}

\end{table}

The \texttt{ S3 summary()} method returns an output that provides an overview of the hyperparameters used in the selection process, such as the measure $\omega$, $k_{\text{max}}$, the copula, and the margins. It then returns a ranking for each survival, reporting the selection frequency.

The function \texttt{PlotBRBVS()} takes as input an object of the class BRBVS and plots the active features against their relative frequencies for the two times to event.

\clearpage
\begin{verbatim}
R> summary(Bivrbvs)

Sets of Relevant Covariates
================================

Metric: FIM 
kmax: 10
Copula: PL 
Margins: PO PO 

================================

Survival Function  1 :
  -  1nd: SevScale1E (53.00%)
  -  2rd: SevScale2E (93.00%)
  -  3rd: ENROLLAGE  (22.00%)

Survival Function  2 :
  -  1nd: SevScale1E (47.00%)
  -  2rd: SevScale2E (91.00%)


\end{verbatim}

\begin{verbatim}
R> BRBVS::plotBRBVS(Bivrbvs)
\end{verbatim}

\begin{figure}[ht]
\centering
\includegraphics[height=0.5\textheight, keepaspectratio]{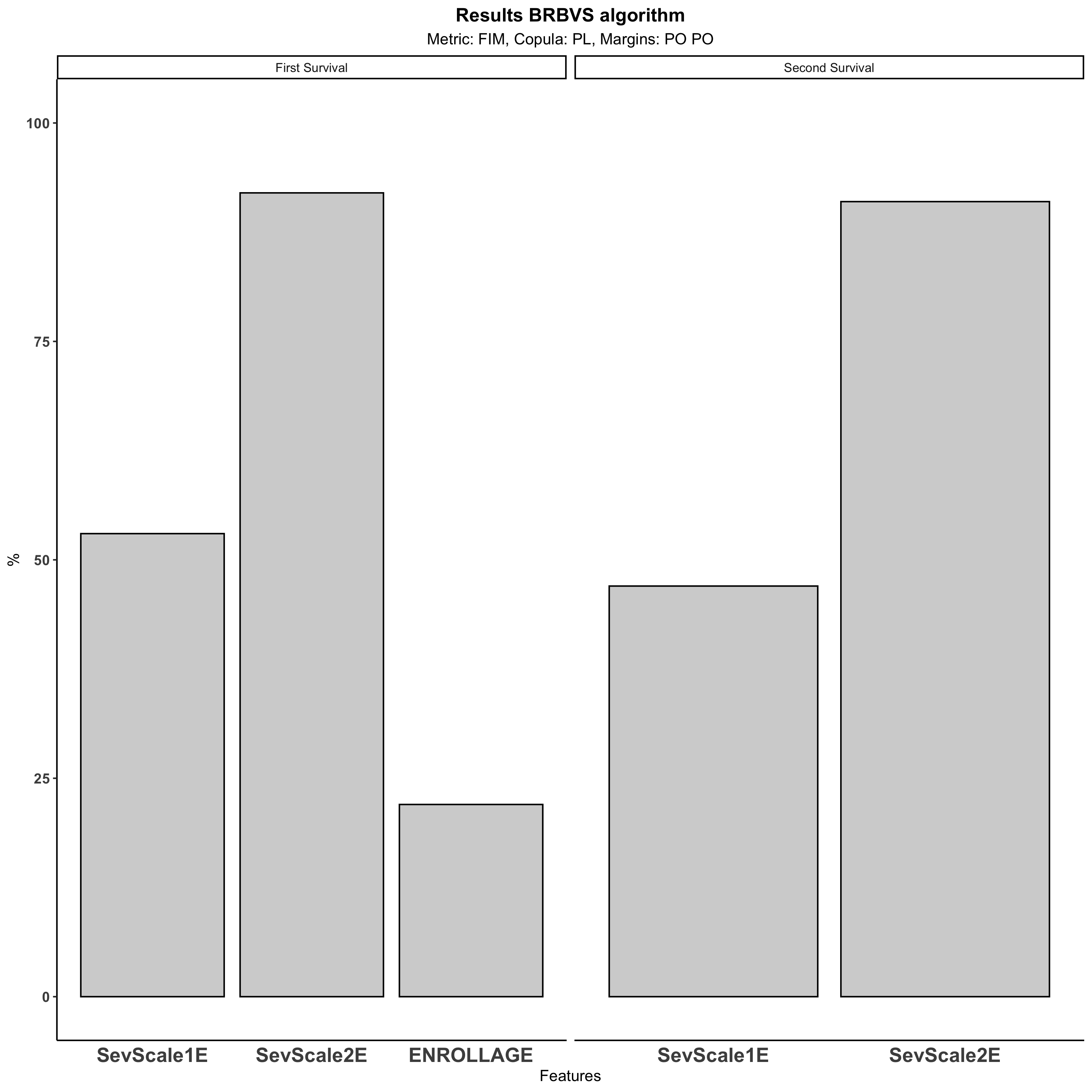}
\caption{Output \texttt{plotBRBVS()} function.}
\label{fig:copulaplot}
\end{figure}

The \texttt{BRBVS} package also implements canonical variable selection methods, such as backward and forward selection, which are well-suited to be adapted to the Copula Link Based Survival model.


In addition to the variable ranking method discussed in Section~\ref{sec:Framework}, the \texttt{BRBVS} package implements variable selection procedures such as backward selection and forward selection. 

The function \texttt{BRBVS::backward\_selection\_BivCop()} performs backward selection for bivariate copula survival models based on Akaike information criterion (\texttt{AIC}) or Bayesian information criterion (\texttt{BIC}). It iteratively removes variables from the model to minimize the specified measure, providing an optimized model with the most relevant variables for each survival function.

The main parameters for \texttt{BRBVS::backward\_selection\_BivCop()} are:
\begin{itemize}
    \item \textbf{\texttt{data}}: A data frame containing \texttt{t11}, \texttt{t12}, \texttt{t21}, \texttt{t22}, \texttt{cens1}, \texttt{cens2}, and the $p$ covariates.
    \item \texttt{lowerBt1} and \texttt{lowerBt2}: Names of the lower bounds for the first and second times-to-event variables (default: \texttt{t11} and \texttt{t21}).
    \item \texttt{upperBt1} and \texttt{upperBt2}: Names of the upper bounds for the first and second times-to-event variables (default: \texttt{t12} and \texttt{t22}).
    \item \textbf{\texttt{copula}}: Specifies the type of copula to be used in the model (default: \texttt{N} for Normal copula).
    \item \texttt{margins}: Character vector defining the margins for the copula model (default: \texttt{c("PH", "PH")}, representing Proportional Hazards).
    \item \texttt{measure}: Criterion to minimize during selection (\texttt{AIC} by default; alternatively, \texttt{BIC}).
    \item \texttt{cens1} and \texttt{cens2}: Censoring indicators for the first and second times-to-event variables.
\end{itemize}

For more details, the arguments of the \texttt{BRBVS::backward\_selection\_BivCop()} function are listed in Table~\ref{tab:infoForwardBackward}, and the algorithm is outlined in Algorithm~\ref{alg:backward}.

\begin{table}[ht]
\renewcommand{\arraystretch}{1.3}
\begin{center}
{\footnotesize
\begin{tabular}{l p{10cm}}
\hline
{Arguments} & {Description} \\
\hline
\texttt{data} & A data frame containing the times to event\texttt{t11,t12,t21,t22}, censoring variables \texttt{cens1, cens2}, and covariates. \\
\texttt{lowerBt1} & Character. Name of the lower bound for the first time to event. \\
\texttt{lowerBt2} & Character. Name of the lower bound for the second time to event. \\
\texttt{upperBt1} & Character. Name of the upper bound for the first time to event. \\
\texttt{upperBt2} & Character. Name of the upper bound for the second time to event. \\
\texttt{copula} & Character. Type of copula to be used in the model. Default is Normal copula (\texttt{"N"}), see Table \ref{tab:infoBRBVS} for the range of copula functions available. \\
\texttt{margins} & Character vector. Margins to be used in the copula model. Default is \texttt{c('PH', 'PH')}. \\
\texttt{measure} & Character. Measure to be minimized during the selection process. Either the Akaike information criterion \texttt{("AIC")} or Bayesian information criterion \texttt{"BIC"}. Default is \texttt{"AIC"}. \\
\texttt{cens1} & Character. Censoring indicator for the first time to event. \\
\texttt{cens2} & Character. Censoring indicator for the second time to event. \\
\hline
\end{tabular}
}
\end{center}
\caption{\texttt{BRBVS::backward\_selection\_BivCop()} and \texttt{BRBVS::forward\_selection\_BivCop()} argument descriptions.}
\label{tab:infoForwardBackward}

\end{table}

\clearpage

\begin{verbatim}
R> data(AREDS)
R> BRBVS::AREDS$GG<- ifelse(AREDS$rs2284665==0,1,0)
R> BRBVS::AREDS$GT<- ifelse(AREDS$rs2284665==1,1,0)
R> BRBVS::AREDS<- AREDS[,!(colnames(AREDS) %in% 'rs2284665')]

R> bs_AREDS<- BRBVS::backward_selection_BivCop(AREDS, 
                                    lowerBt1 = "t11", lowerBt2 = "t21",
                                     upperBt1 = "t12", upperBt2 = "t22",
                                     copula = "PL",
                                     margins = c("PO", "PO"),
                                     measure = "AIC",
                                     cens1, cens2)
        
\end{verbatim}

\begin{verbatim}
R> print(bs_AREDS)

$Results
  Step        Model      AIC
1    1 (full model) 4217.215
2    2   Remove: GT 4214.076

$Equations
$Equations[[1]]
t11 ~ s(t11, bs = "mpi") + SevScale1E + ENROLLAGE + SevScale2E + GG 
<environment: 0x7fb0f993cbd0>

$Equations[[2]]
t21 ~ s(t21, bs = "mpi") + SevScale1E + ENROLLAGE + SevScale2E + GG 
<environment: 0x7fb0f993cbd0>

$Equations[[3]]
~SevScale1E + ENROLLAGE + SevScale2E + GG 
<environment: 0x7fb0f993cbd0>
  
\end{verbatim}

\begin{algorithm}
{\footnotesize
 \caption{Backward Selection  Algorithm implemented in \texttt{BRBVS}.}\label{alg:backward}
 \SetAlgoLined
 {{\bf Input:} $BackwardSelection(\mathbf{Z}, C, g, IC)$}
 \BlankLine
 \# $\mathbf{Z}:[n \times (p+2)]$ data matrix with $\mathbf{z}_i=(t_{1i}, t_{2i}, x_{i1}, \dots, x_{ip})$;\\
 \# $C:$ A copula function among the ones presented in Table \ref{tab:copulafamilies};\\
 \# $g:$ A set of link functions among ones presented in Table \ref{tab:marl};\\
 \# $IC:$ Either AIC or BIC;\\
 \BlankLine
 {{\bf Output:} Optimal set of predictors}
 \BlankLine
 \# Initialize the current model $M_0$ with all predictors;\\
 $\quad M_0 \gets \text{GJRM}(\mathbf{Z})$;\\
 $\quad IC_{0} \gets IC(M_0)$;\\
 $\quad \mathbf{X} \gets \text{Set of all predictors }$;\\
 \BlankLine
 \While{$\mathbf{X}$ is not empty}{
    $IC_{\text{min}} \gets \infty$;\\
    $worst\_predictor \gets \emptyset$;\\
    \BlankLine
    \For{$j \gets 1$ \KwTo $p$}{
        \# Fit model $M_{-x}$ by removing predictor $x$ from $M_0$;\\
        $M_{-j} \gets \text{GJRM}(\mathbf{Z}[,-x_j], C, g)$;\\
        $IC_{-j} \gets IC(M_{-j})$;\\
        \BlankLine
        \If{$IC_{-j} < IC_{\text{min}}$}{
            $IC_{\text{min}} \gets IC_{-j}$;\\
            $worst\_predictor \gets x_j$;\\
        }
    }
    \BlankLine
    \If{$IC_{\text{min}} < IC_{0}$}{
        \# Remove $worst\_predictor$ from $M_0$;\\
        $M_0 \gets \text{Model with } worst\_predictor \text{ removed}$;\\
        $IC_{0} \gets IC_{\text{min}}$;\\
        $\mathbf{X} \gets \mathbf{X} \setminus \{worst\_predictor\}$;\\
    }
    \Else{
        \textbf{break};\\
    }
 }
 \BlankLine
 \textbf{return} $M_0$;\\
 }
\end{algorithm}

\clearpage

The \texttt{BRBVS::forward\_selection\_BivCop()} function performs forward selection for bivariate copula survival models, based on Akaike information criterion (\texttt{AIC}) or Bayesian information criterion (\texttt{BIC}). It iteratively adds variables to the model to minimize the specified measure.

The key arguments are:
\begin{itemize}
    \item {\texttt{data}}: A data frame containing \texttt{t11}, \texttt{t12}, \texttt{t21}, \texttt{t22}, \texttt{cens1}, \texttt{cens2}, and the $p$ covariates.
    \item {\texttt{lowerBt1} and \texttt{lowerBt2}}: Names of the lower bounds for the first and second times-to-event variables (default: \texttt{t11} and \texttt{t21}).
    \item {\texttt{upperBt1} and \texttt{upperBt2}}: Names of the upper bounds for the first and second times-to-event variables (default: \texttt{t12} and \texttt{t22}).
    \item {\texttt{copula}}: Specifies the type of copula to be used in the model (default: \texttt{N} for Normal copula).
    \item {\texttt{margins}}: Character vector defining the margins for the copula model (default: \texttt{c("PH", "PH")}, representing Proportional Hazards).
    \item {\texttt{measure}}: Criterion to minimize during selection (\texttt{AIC} by default; alternatively, \texttt{BIC}).
    \item {\texttt{cens1} and \texttt{cens2}}: Censoring indicators for the first and second times-to-event variables.
\end{itemize}

Further details about the arguments of the \texttt{BRBVS::forward\_selection\_BivCop()} function are provided in Table~\ref{tab:infoForwardBackward}, and the algorithm is outlined in Algorithm~\ref{alg:forward}.

\begin{verbatim}
R> fs_AREDS<- BRBVS::forward_selection_BivCop(data = AREDS,
                        lowerBt1 = "t11", lowerBt2 = "t21",
                        upperBt1 = "t12", upperBt2 = "t22", 
                        copula = "N", 
                        margins = c("PH", "PH"), 
                        measure = "AIC", cens1, cens2 )
\end{verbatim}

\begin{verbatim}
R> print(fs_AREDS)

$Results
  Step       Model      AIC
1    1 (intercept) 4488.600
2    2  SevScale2E 4270.011
3    3  SevScale1E 4222.604
4    4          GG 4218.154
5    5   ENROLLAGE 4214.076

$Equations
$Equations[[1]]
t11 ~ s(t11, bs = "mpi") + SevScale2E + SevScale1E + GG + ENROLLAGE
<environment: 0x7fb0f6251568>

$Equations[[2]]
t21 ~ s(t21, bs = "mpi") + SevScale2E + SevScale1E + GG + ENROLLAGE
<environment: 0x7fb0f6251568>

$Equations[[3]]
~SevScale2E + SevScale1E + GG + ENROLLAGE
<environment: 0x7fb0f6251568>
\end{verbatim}

The output, in addition to displaying the table with the steps and the value of the chosen measure, also provides the three optimal equations based on the forward selection.

\begin{algorithm}
{\footnotesize
 \caption{Forward Selection Algorithm implemented in \texttt{BRBVS}.}\label{alg:forward}
 \SetAlgoLined
 {{\bf Input:} $ForwardSelection(\mathbf{Z}, C, g, IC)$}
 \BlankLine
 \# $\mathbf{Z}:[n \times 2]$ data matrix with $\mathbf{z}_i=(t_{1i}, t_{2i})$;\\
 \# $C:$ A copula function among the ones presented in Table \ref{tab:copulafamilies};\\
 \# $g:$ A set of link functions  presented in Table \ref{tab:marl};\\
 \# $IC:$ Either AIC or BIC;\\
 \BlankLine
 {{\bf Output:} Optimal set of predictors}
 \BlankLine
 \# Initialize the current model $M_0$ with no predictors;\\
 $\quad M_0 \gets \text{GJRM}(\mathbf{Z})$;\\
 $\quad IC_{\text{0}} \gets IC(M_0)$;\\
 $\quad \mathbf{X} \gets \text{Set of all predictors}$;\\
 \BlankLine
 \While{$\mathbf{X}$ is not empty}{
    $IC_{\text{min}} \gets \infty$;\\
    $best\_predictor \gets \emptyset$;\\
    \BlankLine
    \For{each predictor $x \in P$}{
        \# Fit model $M_x$ by adding predictor $x$ to $M_0$;\\
        $M_j \gets \text{GJRM}(\mathbf{Z}, x_j, C, g)$;\\
        $IC_j \gets IC(M_j)$;\\
        \BlankLine
        \If{$IC_j < IC_{\text{min}}$}{
            $IC_{\text{min}} \gets IC_j$;\\
            $best\_predictor \gets x_j$;\\
        }
    }
    \BlankLine
    \If{$IC_{\text{min}} < IC_{0}$}{
        \# Update $M_0$ to include $best\_predictor$;\\
        $M_0 \gets \text{Model with } best\_predictor \text{ added}$;\\
        $IC_{0} \gets IC_{\text{min}}$;\\
        $\mathbf{X} \gets \mathbf{X} \setminus \{best\_predictor\}$;\\
    }
    \Else{
        \textbf{break};\\
    }
 }
 \BlankLine
 \textbf{return} $M_0$;\\
 }
\end{algorithm}

\clearpage
\subsubsection{Copula Link Based Survival Model estimation through GJRM package}

To make the paper self contained, we here describe how the selected variables can then considered in the \texttt{GJRM} package \citep{Man:GJRM} to estimate the Copula Link Survival Model.

Once we have the two sets of relevant variables, we can estimate the model through the \texttt{GJRM::gjrm()} function in the \texttt{GJRM} package  in \texttt{R} and obtain the parameter estimates.

The first step is to specify the three linear predictors. In this case, we found that \texttt{ENROLLAGE, SevScale1E, SevScale2E} are relevant covariates for the first margin, while \texttt{SevScale1E, SevScale2E} are relevant for the second margin, this based on the BRBVS algorithm. For the third margin, which pertains to the dependence parameter, no variable selection was performed. In this instance, we include \texttt{SevScale1E, SevScale2E} since they are the intersection of the two sets obtained.

\begin{verbatim}
R> eta1 <- t11 ~ s(t11, bs = "mpi") + s(ENROLLAGE) + SevScale1E + SevScale2E
R> eta2 <- t21 ~ s(t21, bs = "mpi")  + SevScale1E + SevScale2E
R> eta3 <- ~ SevScale1E + SevScale2E
\end{verbatim}

 we are specifying \(\eta_1\) (\texttt{eta1}), \(\eta_2\) (\texttt{eta2}), and \(\eta_3\) (\texttt{eta3}). Additionally, we note that a spline \texttt{s()} has been applied to \texttt{ENROLLAGE} since we observe that the progression of the disease has a non-linear effect with age.

\begin{verbatim}
R> f.list <- list(eta1, eta2, eta3)

R> out <- GJRM::gjrm(f.list, data = AREDS, surv = TRUE,
            copula = "PL", margins = c("PO", "PO"),
            cens1 = cens1, cens2 = cens2, model = "B",
            upperBt1 = 't12', upperBt2 = 't22')   
\end{verbatim}

Setting \texttt{surv=TRUE} indicates that a bivariate survival model is being fitted. The \texttt{copula} parameter is specified as \texttt{"PL"}, indicating the use of a Plackett copula. The \texttt{margins} parameter is set to \texttt{c("PO", "PO")}, where \texttt{"PO"} refers to the proportional odds model. The \texttt{cens1} and \texttt{cens2} parameters provide the censoring indicators for the two survival variables. The \texttt{model='B'} denotes the application of a bivariate model. Lastly, the \texttt{upperBt1} and \texttt{upperBt2} parameters define the upper bounds for the two times to event variables.

The model estimation is based on a two-stage process using the trust region algorithm; for details about the model framework and the implementation, see \cite{Marra2020, Man:GJRM}. Therefore, it is always a good practice to verify that the convergence has been correctly achieved, ensuring that the gradient values are acceptable and that the information matrix is positive definite. We can then check the convergence of the model using the function. \texttt{conv.check()} in \texttt{GJRM} package.

\begin{verbatim}
R> GJRM::conv.check(out)

Largest absolute gradient value: 4.166472e-05
Observed information matrix is positive definite
Eigenvalue range: [0.008964266,164318.3]

Trust region iterations before smoothing parameter estimation: 71
Loops for smoothing parameter estimation: 8
Trust region iterations within smoothing loops: 18
Estimated overall probability range: 0.02390308 0.9999404
Estimated overall density range: 5.964254e-05 7.811673

  \end{verbatim}

The model successfully converged, as evidenced by the number of iterations completed by the trust region algorithm (71). Additionally, the probability values obtained are coherent [0.02390308,  0.9999404]. Furthermore, the positively defined nature of the observed information matrix reaffirms the reliability of our results, eigenvalues range [5.964254e-05, 7.811673].

\begin{verbatim}
    R> summary(out)

COPULA:   Plackett
MARGIN 1: survival with -logit link
MARGIN 2: survival with -logit link

EQUATION 1
Formula: t11 ~ s(t11, bs = "mpi") + s(ENROLLAGE) + SevScale1E + SevScale2E

Parametric coefficients:
            Estimate Std. Error z value Pr(>|z|)    
(Intercept) -18.3487     4.3956  -4.174 2.99e-05 ***
SevScale1E5   0.6875     0.2645   2.599 0.009355 ** 
SevScale1E6   0.8084     0.2552   3.168 0.001537 ** 
SevScale1E7   1.7219     0.2750   6.261 3.81e-10 ***
SevScale1E8   2.5884     0.3608   7.175 7.25e-13 ***
SevScale2E5   0.4063     0.2813   1.444 0.148612    
SevScale2E6   0.8768     0.2692   3.258 0.001123 ** 
SevScale2E7   1.0164     0.2900   3.505 0.000457 ***
SevScale2E8   1.4862     0.3403   4.367 1.26e-05 ***
---
Signif. codes:  0 ‘***’ 0.001 ‘**’ 0.01 ‘*’ 0.05 ‘.’ 0.1 ‘ ’ 1

Smooth components' approximate significance:
               edf Ref.df   Chi.sq p-value    
s(t11)       6.649  7.674 1836.475  <2e-16 ***
s(ENROLLAGE) 1.625  2.039    5.807  0.0548 .  
---
Signif. codes:  0 ‘***’ 0.001 ‘**’ 0.01 ‘*’ 0.05 ‘.’ 0.1 ‘ ’ 1


EQUATION 2
Formula: t21 ~ s(t21, bs = "mpi") + SevScale1E + SevScale2E

Parametric coefficients:
            Estimate Std. Error z value Pr(>|z|)    
(Intercept) -30.4468    10.5514  -2.886 0.003907 ** 
SevScale1E5   0.2684     0.2572   1.043 0.296794    
SevScale1E6   0.3454     0.2459   1.405 0.160059    
SevScale1E7   0.8530     0.2592   3.291 0.000999 ***
SevScale1E8   0.9947     0.3286   3.027 0.002468 ** 
SevScale2E5   0.8682     0.2796   3.105 0.001903 ** 
SevScale2E6   1.2294     0.2747   4.475 7.64e-06 ***
SevScale2E7   2.2515     0.2988   7.534 4.92e-14 ***
SevScale2E8   3.4251     0.3632   9.431  < 2e-16 ***
---
Signif. codes:  0 ‘***’ 0.001 ‘**’ 0.01 ‘*’ 0.05 ‘.’ 0.1 ‘ ’ 1

Smooth components' approximate significance:
         edf Ref.df Chi.sq p-value    
s(t21) 7.374  8.204   3808  <2e-16 ***
---
Signif. codes:  0 ‘***’ 0.001 ‘**’ 0.01 ‘*’ 0.05 ‘.’ 0.1 ‘ ’ 1


EQUATION 3
Link function for theta: log 
Formula: ~SevScale1E + SevScale2E

Parametric coefficients:
            Estimate Std. Error z value Pr(>|z|)    
(Intercept)   0.9658     0.5089   1.898 0.057702 .  
SevScale1E5  -0.2302     0.5196  -0.443 0.657677    
SevScale1E6  -0.6698     0.4964  -1.349 0.177259    
SevScale1E7  -1.1222     0.5165  -2.173 0.029812 *  
SevScale1E8  -1.0086     0.6035  -1.671 0.094662 .  
SevScale2E5   1.2973     0.5464   2.374 0.017580 *  
SevScale2E6   1.9442     0.5167   3.763 0.000168 ***
SevScale2E7   1.7939     0.5365   3.344 0.000826 ***
SevScale2E8   1.1878     0.6251   1.900 0.057421 .  
---
Signif. codes:  0 ‘***’ 0.001 ‘**’ 0.01 ‘*’ 0.05 ‘.’ 0.1 ‘ ’ 1

theta = 6.44(3.13,13.8)  tau = 0.354(0.203,0.496)
n = 628  total edf = 42.6
\end{verbatim}

Our analysis involves three distinct margin specifications, and correspondingly, the \texttt{summary()} function returns three tables, each dedicated to one margin. These output tables are organized into two key sections: `Parametric Effects' and `Smooth Effects'. In the `Parametric Effects' section,  the estimates alongside their associated standard errors, including the z-value and the corresponding p-values are presented. The `Smooth Effects' section focuses on the complexity of the model's smooth components. This is detailed by the number of degrees of freedom (\texttt{edf}), where, for instance, a value of `1' suggests a simple straight line, while `2' suggests a curve. Additionally, \texttt{Ref.df} (Reference degrees of freedom) and \texttt{chi-square} values are utilized to assess the statistical significance of these smooth components. Finally,  we have the value of \texttt{theta} which represents $\theta_i= m\left\{\eta_{3 i}\left(\mathbf{x}_{3 i} ; \boldsymbol{\beta}_{3}\right)\right\}$,  \texttt{tau} denotes the estimate of Kendall's tau, the sample size \texttt{n} and the estimate of the degrees of freedom  \texttt{total.edf} for the model specified. The functions \texttt{AIC()} and \texttt{BIC()} work in a similar fashion as those for the classical statistical models. More details about the model output produced by the function \texttt{GJRM::gjrm()} can be found in the \texttt{GJRM} package in \texttt{R}.

\clearpage
\section{Simulation Study}
\label{sec:simulation}

This section explores the R functions used to develop the simulation study presented in \cite{Petti2024sub} and  \cite{PETTI22}, introducing, for the first time, methods to simulate bivariate copula survival structures with censoring. Additionally, we highlight key results from the simulation study in \cite{Petti2024sub}.

Let $\mathbf{X}\in \mathbb{R}^{n \times p}$ be the design matrix partitioned as $\mathbf{X}=(\mathbf{X}_{11} : \mathbf{X}_{12})$, where $\mathbf{X}_{11}$ contains the \emph{informative covariates}, 
$\mathbf{X}_{11}=(\boldsymbol{x}_1, \boldsymbol{x}_2, \boldsymbol{x}_3) \in \mathbb{R}^{n \times 3}$, with $\boldsymbol{x}_1, \boldsymbol{x}_2, \boldsymbol{x}_3$ three $n$-dimensional vectors, while the remaining $p-3$ variables included in $\mathbf{X}_{12}$ are  \emph{non-informative}.
In more detail, $\mathbf{X}_{11}$ and $\mathbf{X}_{12}$  are generated from a multivariate Gaussian distribution such that 
$\mathbf{X}_{11} \sim \mathcal{N}_3(\mathbf{0}, \boldsymbol{\Sigma}_{\mathbf{X}{11}})$, with $\mathbf{0}$ a null vector of means, and $\boldsymbol{\Sigma}_{\mathbf{X}_{11}}$ is the covariance matrix with diagonal elements fixed to one and off-diagonal elements equal to $0.5$;  $\mathbf{X}_{12} \sim \mathcal{N}_{p-3}(\mathbf{0}, \boldsymbol{\Sigma}_{\mathbf{X}_{12}})$  with $\boldsymbol{\Sigma}_{\mathbf{X}_{12}}$ an identity matrix.

 The two times to event $T_{\nu 1}$ and $T_{\nu 2}$ were generated from a Proportional Hazards and Odds model, respectively. In \texttt{R} code this can be breakdown as follows

\begin{verbatim}

baseline_survival <- function(t) {
    0.9 * exp(-0.4 * t^2.5) + 0.1 * exp(-0.1 * t^1)
}

f1 <- function(t, beta1, beta2, u, z1, z2) {
    S_0 <- baseline_survival(t)
    transformed_value <- exp(-exp(log(-log(S_0)) 
                                  + beta1 * z1 + beta2 * z2))
    result <- transformed_value - u
    return(result)
}

f2 <- function(t, beta1, beta2, u, z1, z3) {
    S_0 <- baseline_survival(t)
    transformed_value <- 1 / (1 + exp(log((1 - S_0) / S_0) 
                                      + beta1 * z1 + beta2 * z3))
    result <- transformed_value - u
    return(result)
}
    
\end{verbatim}

where we first define the baseline survival as $S(t)= 0.9  \exp(-0.4  t^{2.5}) + 0.1 \exp(-0.1 t^1) $ and then the transformed values are obtained using the formulae in Table \ref{tab:marl}. The details relating to the survival functions, the baseline and the parameters used for the generation of the times to event are summarized in the Table \ref{tab:SimMargins}

\begin{table}[h]
\renewcommand{\arraystretch}{1.3}
\begin{center}
{\footnotesize 
\begin{tabular}{c|ll}
\hline
\textbf{Parameter} & \textbf{Model for $T_{1i}$} & \textbf{Model for $T_{2i}$} \\
\hline
Model Type & Proportional Hazards (\texttt{PH}) & Proportional Odds (\texttt{PO}) \\
\hline
Formula & $T_{1i}=\log[-\log{S_{10}(t_{1i})}]+\beta_{11}x_{1i}+\beta_{12}x_{2i}$ & $T_{2i}=\log\left[\frac{(1-S_{20}(t_{2i}))}{S_{20}(t_{2i})}\right]+\beta_{21}x_{1i}+\beta_{22}x_{3i}$ \\
$S(t)$ & $S_{10}(t_{1i})=0.9e^{-0.4t_{1i}^{2.5}}+ 0.1 e^{-0.1 t_{1i}}$ & $S_{20}(t_{2i})=0.9e^{-0.4t_{2i}^{2.5}} + 0.1 e^{-0.1 t_{2i}}$ \\
$\beta_{11}$ & $-1.5$ & - \\
$\beta_{12}$ & $1.7$ & - \\
$\beta_{21}$ & - & $-1.5$ \\
$\beta_{22}$ & - & $-1.3$ \\
\hline
\end{tabular}
}
\end{center}
\caption{Margins generation process with $S(\cdot)$ the survival function, $\log(\cdot)$ the natural logarithm, and $e(\cdot)$ the corresponding base.}
\label{tab:SimMargins}

\end{table}

The times are generated with Brent's univariate root-finding method. Practically we use the \texttt{unitroot()} function in \texttt{R}, specifying a range between $0$ and $8$, and allowing for potential extensions. The random censoring times are derived from the lower and upper bounds of two independent uniform random variables. These bounds are then compared with the simulated times to assign censoring.

\begin{verbatim}
u <- stats::runif(n, 0, 1)
t <- rep(NA, n)

for (i in 1:n) {
    t[i] <- stats::uniroot(f1, c(0, 8), tol = .Machine$double.eps^0.5,
                           beta1 = beta11, beta2 = beta12, u = u[i],
                           z1 = z1[i], z2 = z2[i], extendInt = "yes")$root
}

c1 <- stats::runif(n, 0, 2)
c2 <- c1 + stats::runif(n, 0, 6)

dataSim <- data.frame(t.true1 = t, c11 = c1, c12 = c2, t11 = NA,
                      t12 = NA, z1, z2, z3, cens = character(n),
                      surv1 = u, stringsAsFactors = FALSE)

for (i in 1:n) {
    if (t[i] > c2[i]) {
        dataSim$t11[i] <- c2[i]
        dataSim$t12[i] <- NA 
        dataSim$cens[i] <- "R"
    } else {
        dataSim$t11[i] <- t[i]
        dataSim$t12[i] <- NA
        dataSim$cens[i] <- "U"
    }
}
\end{verbatim}

Before simulating $T_2$, it is useful to specify the degree of dependence between the two times to event variables. The two main components in the recipe are the dependence parameter $\theta$ and the copula function as in Table \ref{tab:copulafamilies}, Clayton in our example. This framework ensures the generation of $T_2$ observations that are dependent on the already observed values for $T_1$.

\begin{verbatim}
eta.theta <- 1.2
theta <- exp(eta.theta)

u2 <- stats::runif(n, 0, 1)
u_prime <- ((u2^(-theta/(1 + theta)) - 1) * u^(-theta) + 1)^(-1/theta)
t <- rep(NA, n)

for (i in 1:n) {
    t[i] <- stats::uniroot(f2, c(0, 8), tol = .Machine$double.eps^0.5,
                           beta1 = beta21, beta2 = beta22, u = u_prime[i],
                           z1 = z1[i], z3 = z3[i], extendInt = "yes")$root
}

dataSim$t.true2 <- t
c1 <- stats::runif(n, 0, 2)
c2 <- c1 + stats::runif(n, 0, 6)
dataSim$c21 <- c1
dataSim$c22 <- c2

for (i in 1:n) {
    if (t[i] > c2[i]) {
        dataSim$t21[i] <- c2[i]
        dataSim$t22[i] <- NA
        dataSim$cens[i] <- paste(dataSim$cens[i], "R", sep = "")
    } else {
        dataSim$t21[i] <- t[i]
        dataSim$t22[i] <- NA
        dataSim$cens[i] <- paste(substr(dataSim$cens[i], 1, 1), 
                                 "U", sep = "")
    }
}
\end{verbatim}

This implies that the covariate $\boldsymbol{x}_2$, although not appearing in the formula on the right in Table \ref{tab:SimMargins}, is certainly a relevant feature for $T_2$. From a causal inference perspective, the set of covariates involved in the data generating process of $T_2$ is $\{\boldsymbol{x}_1, \boldsymbol{x}_2, \boldsymbol{x}_3\}$. We consider two different scenarios of dependence between the two survival margins:

\begin{itemize}
    \item \textbf{Scenario A}: $\eta_{3i}=\beta_{30}$;
    \item \textbf{Scenario B}: $\eta_{3i}=\beta_{31}x_{1i}+\beta_{32}x_{2i}+\beta_{33}x_{3i}$,
\end{itemize}

with $\boldsymbol{\beta}_3=(\beta_{30}, \beta_{31}, \beta_{32}, \beta_{33})=(1.2, -1.5, 1.7, -1.5)$. The values of $\eta_{3i}$ are specified to range the Kendall tau value between $0.10$ and $0.90$. It is important to note that $\eta_{3i}$ represents the potential variability in the dependence between $(T_{1i}, T_{2i})$ across different observations.
Further, in Scenario A, we are considering weak dependence between $(T_{1i}, T_{2i})$, while in Scenario B, the data are generated by considering a dependence between $(T_{1i}, T_{2i})$.

Moreover, the data points experience right-censoring, resulting in $11\%$ and $32\%$ missing information for the first and second times, respectively. Specifically, random censoring times are derived from the lower and upper bounds of two independent uniform random variables. These bounds are then compared with the simulated times to assign censoring.

The simulation study is based on $n_{\text{rep}}=100$ Monte Carlo replicates and for each run we consider $B=50$ bootstrap replicates. 
The threshold parameter $\tau$ in \eqref{eq:hat_s_v} is set at $\tau=0.5$, whereas the maximum number of important covariates is $k_{\text{max}}=6$.

In the model class discussed in Section \ref{Ch:modelformulation}, the sample size significantly influences the precision of the estimates, as highlighted in Supplementary Material G in \cite{PETTI22}. To achieve this precision, we select the two sample sizes  $n=\{800, 1000\}$, to allow bootstrap samples with an appropriate number of units. For this aim, we fix $r=2$ (as in \cite{Baranowski2020}) and consequently  the bootstrap subsets contain $m=\{400, 500\}$ units, respectively.


The measures used in the simulation study to rank the variables are two:
(a) $\hat{\omega}_{\nu j}$ based on the Fisher information matrix;
(b) the absolute value of the estimated coefficients $|\hat{\beta}_{\nu_j}|$, denoted by $\hat\phi_{\nu j}$, for $(\nu=1,2; j=1,\dots, p)$.

This last measure is largely used in the regression context to screen variables even in the presence of datasets of large dimension (see among the others \cite{FanSong2010}) but, differently from the proposed measure (a), it considers only the punctual value of the estimate both discarding other information that can be taken from the likelihood and neglecting the intrinsic dependence of the copula function.

The performance of the BRBVS  obtained through the \texttt{BRBVS()} function in \texttt{R} \citep{Petti2024Package} is assessed by computing different metrics based on the evaluation of False Positive (FP) and False Negative (FN), where FP is the number of covariates incorrectly chosen as relevant by the variable selection procedure, and  FN is the number of covariates incorrectly chosen as irrelevant by the variable selection procedure. Then, we compute the average of FP and of FN across all $n_{\text{rep}}$ replicates. Specifically, given the average of FP  is equal to $FP_{\nu}=n_{\text{rep}}^{-1}\sum_{h=1}^{n_{\text{rep}}} FP_{\nu}^{(h)}$ and the average of FN is $FN_{\nu}=n_{\text{rep}}^{-1}\sum_{h=1}^{n_{\text{rep}}} FN_{\nu}^{(h)}$, $FP_{\nu}^{(h)}$ and $FN_{\nu}^{(h)}$ denote the number of false positive and false negative, respectively, for the $\nu^{\text{th}}$ survival outcome in the $h^{\text{th}}$ replicate, for $\nu=1,2$.

Additionally, the average number of variables selected in the estimated relevant set for the $\nu^{\text{th}}$ survival is represented by $\langle\hat{{s}}_\nu\rangle=n_{\text{rep}}^{-1} \sum_{h=1}^{n_{\text{rep}}} |\hat{s}^{(h)}_{\nu}|$, where $|\hat{s}^{(h)}_{\nu}|$ is the estimated size of relevant variables in the $h^{\text{th}}$ replicate.

Furthermore, denote with $s_\nu$, for $\nu=1,2$, the true sets of relevant variables for the first and second survival outcomes, where in our study $s_1=\{1,2\}$ and $s_{2}=\{1, 2, 3\}$. The average number of correctly identified relevant variables across replicates is given by $\langle\hat{s}_{\nu}\cap s_\nu\rangle=n_{\text{rep}}^{-1} \sum_{h=1}^{n_{\text{rep}}}|\hat{s}^{(h)}_{\nu}\cap s_\nu|$, where $|\hat{s}^{(h)}_{\nu}\cap s_\nu|$ counts the relevant variables correctly selected in each replicate.

\begin{sidewaysfigure}
\centering
\includegraphics[height=0.7\textheight, keepaspectratio]{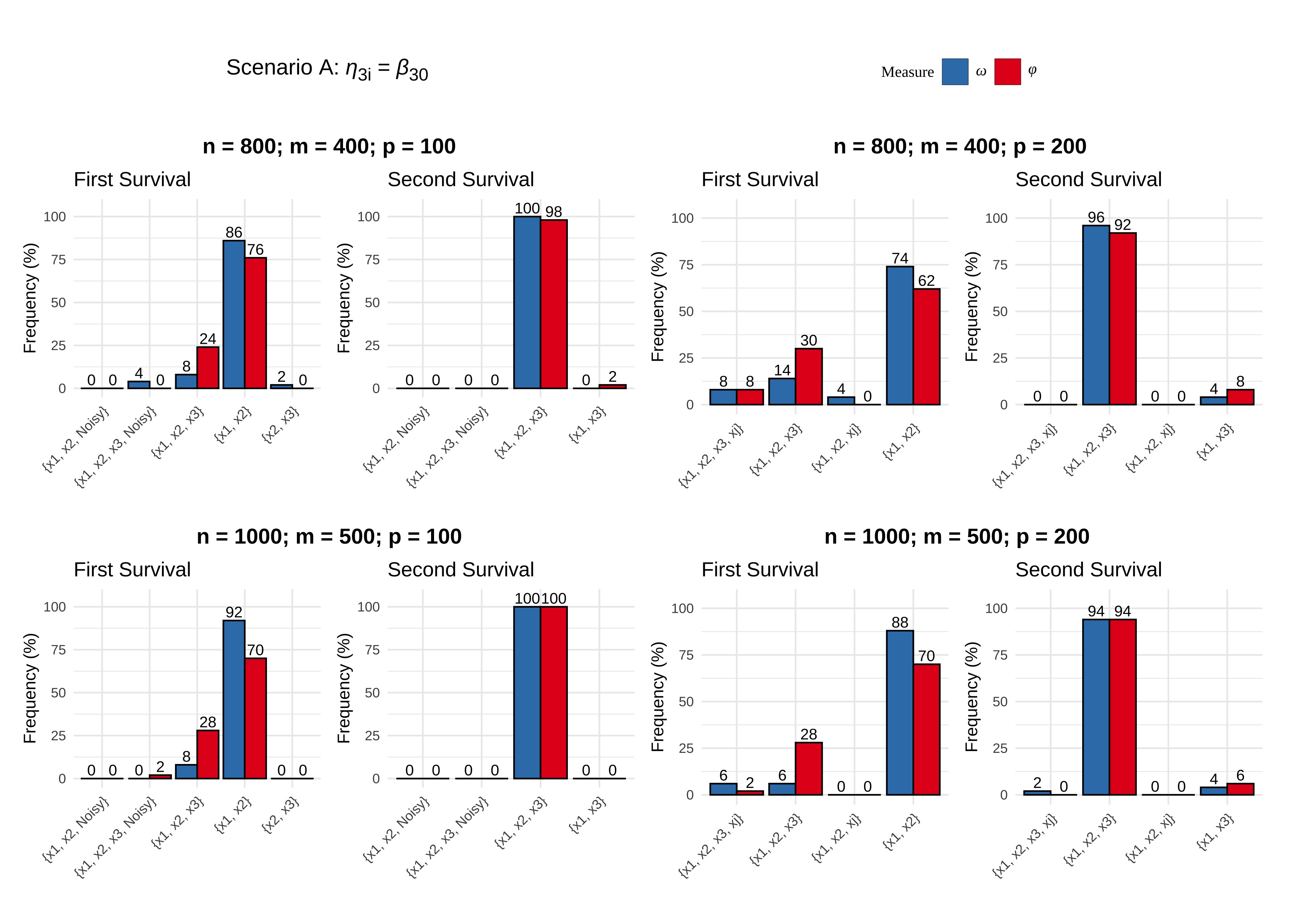}
\caption{Results from simulations highlight the frequency with which relevant sets are selected by the \texttt{BRBVS()} function in \texttt{R}, in the case of Scenario A: $\eta_{3i} = \beta_{30}$.
  }
\label{fig:ScenarioAPlot}
\end{sidewaysfigure}

\begin{sidewaysfigure}
\centering
\includegraphics[height=0.7\textheight, keepaspectratio]{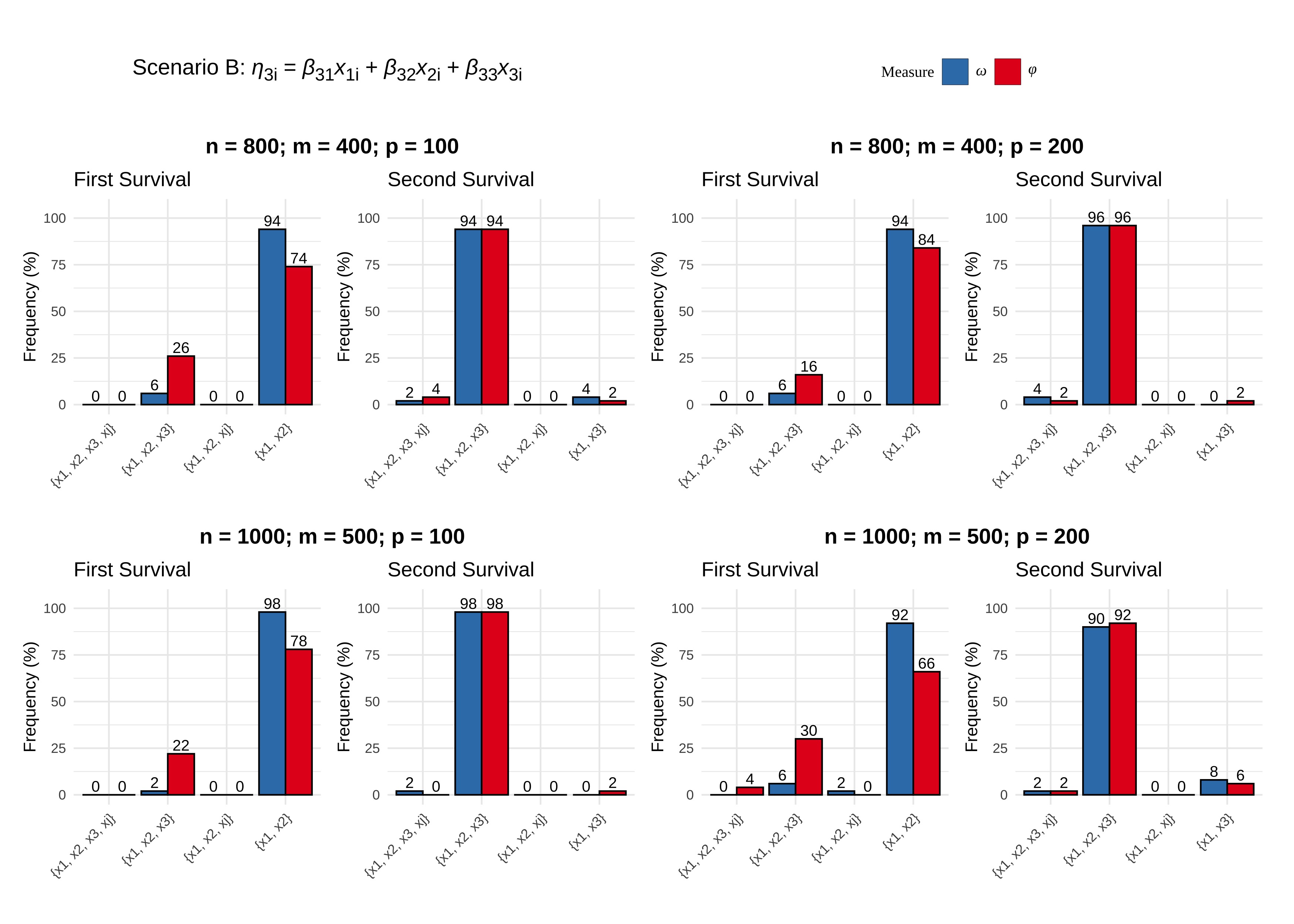}
\caption{Results from simulations highlight the frequency with which relevant sets are selected by the \texttt{BRBVS()} function in \texttt{R}, in the case of Scenario B: $\eta_{3i} = \beta_{31} x_{1i}+\beta_{32} x_{2i}+ \beta_{33}x_{3i}$.
}
\label{fig:ScenarioBPlot}
\end{sidewaysfigure}

For {\em Scenario A} in Figure \ref{fig:ScenarioAPlot}, it can be noted that the metric $\hat\omega$ always selects the two relevant covariates $\boldsymbol{x}_1$ and $\boldsymbol{x}_2$ of the first margin  and the percentage of cases where only $\{\boldsymbol{x}_1, \boldsymbol{x}_2\}$ are chosen increases with $n$.\\
Also in the second margin, all three covariates are always selected when $p=100$, even if this percentage slightly decreases when $p=200$ (as expected).
These last results on the second margin are almost replicated when the $\hat\phi$ metric is considered. \\
For the more complex {\em Scenario B}, the results in Figure \ref{fig:ScenarioBPlot} show that with the $\hat\omega$ metric the covariates $\boldsymbol{x}_1$ and $\boldsymbol{x}_2$ are always selected for the first margin,
and its performance improves as $n$ increases and $p=100$.
For the second margin the three covariates  $\boldsymbol{x}_1$, $\boldsymbol{x}_2$ and $\boldsymbol{x}_3$ are always chosen and they are frequently the only variables included in $\hat s_2$. Comparable results are observed in the case where $p=200$.\\
If we evaluate the performance of the $\hat\phi$ competing metric in  Scenario B, we can further confirm its lower accuracy (with respect to $\hat\omega$) in both cases, when $p=100$  and $p=200$. \\
 All previous results further highlight how the selection of the proper metric, for a given statistical model, can improve the performance of the BRBVS algorithm and consequently the identification of the relevant variables.

\clearpage
\section{Summary} \label{sec:summary}

The \texttt{BRBVS} is the first R package to make variable selection for the class of Bivariate Survival Copula Models \citep{Marra2020}. It extends the Ranking Based Variable Selection algorithm of \cite{Baranowski2020}, originally proposed for linear regression models, to the survival domain considering two main steps: the first ranks and screens the covariates; the second step makes the selection of the relevant variables (for all details see \cite{Petti2024sub}).

The novelties included in the packages are mainly two: 1) the implementation of a procedure that jointly (and not simply marginally) selects the relevant variables of both margins;  2) the implementation of a ranking measure for the covariates that is completely new (as clarified in Section \ref{sec:BRBVS}). The proposed measure is based on the use of the Fisher Information matrix, is almost general and may also be extended to other classes of models. 

To easily allow comparisons with other variable selection approaches, the \texttt{BRBVS} package includes two well established variable selection methods, the forward and backward, where the selection of the relevant variables is based on information criteria (AIC or BIC). 
Finally, the package also includes a function that allows to select the best link for the Bivariate Survival Copula models, among a set of candidate link functions, such that the AIC or BIC is minimized. 

This last feature expands the \texttt{BRBVS} package, originally focused on the variable selection, also to the model selection domain.

\clearpage



\clearpage

\renewcommand{\appendix}{}
\renewcommand{\appendixname}{S}

\renewcommand{\thesection}{S\arabic{section}}
\setcounter{section}{0}

\renewcommand{\thepage}{S\arabic{page}}
\setcounter{page}{1}

\renewcommand{\thetable}{S\arabic{table}}
\setcounter{table}{0}

\renewcommand{\thefigure}{S\arabic{figure}}
\setcounter{figure}{0}

\renewcommand{\theequation}{S\arabic{equation}}
\setcounter{equation}{0}

\renewcommand{\thetheorem}{\arabic{theorem}}
\setcounter{theorem}{1}

\end{document}